\documentclass[%
 reprint,
 amsmath,amssymb,
 aps,
]{revtex4-2}

\usepackage{graphicx}
\usepackage{dcolumn}
\usepackage{bm}
\usepackage{hhline}
\usepackage[caption=false]{subfig}
\usepackage{multirow}
\usepackage{booktabs}
\usepackage{array}
\usepackage{booktabs}
\usepackage{nicefrac}
\newcommand{\ItemSpacing}{0pt}%
\newcommand{\ParSpacing}{0pt}%

\newcommand{\UpperState}{3^{3\!}S_1}%
\newcommand{\MetastableState}{2^{3\!}S_1}%
\newcommand{\WimState}{2^{1\!}S_0}%
\newcommand{\MidState}{2^{3\!}P_{0,1,2}}%
\usepackage{outlines}
\usepackage{enumitem}
\setenumerate[1]{itemsep={\ItemSpacing},parsep={\ParSpacing},label=\Roman*.}
\setenumerate[2]{itemsep={\ItemSpacing},parsep={\ParSpacing},label=\Alph*.}
\setenumerate[3]{itemsep={\ItemSpacing},parsep={\ParSpacing},label=\roman*.}
\setenumerate[4]{itemsep={\ItemSpacing},parsep={\ParSpacing},label=\alph*.}
\usepackage[table]{xcolor}
\definecolor{lightgray}{gray}{0.9}
\usepackage{colortbl}
\usepackage{booktabs}
\definecolor{Gray}{gray}{0.9}

\newcommand{\avg}[1]{\left\langle #1 \right\rangle}
\usepackage{tabularx}
\usepackage{titlesec} 
\usepackage{hyperref}

\begin{document}

\preprint{APS/123-QED}

\title{Direct Measurement of the Forbidden \(\MetastableState \rightarrow \UpperState\) Atomic Transition in Helium}

\author{K. F. Thomas}
\affiliation{Laser Physics Centre, Research School of Physics, The Australian National University,\\ Canberra, ACT 2601, Australia}

\author{J. A. Ross}
\affiliation{Laser Physics Centre, Research School of Physics, The Australian National University,\\ Canberra, ACT 2601, Australia}

\author{B. M. Henson}
\affiliation{Laser Physics Centre, Research School of Physics, The Australian National University,\\ Canberra, ACT 2601, Australia}

\author{D. K. Shin}
\affiliation{Laser Physics Centre, Research School of Physics, The Australian National University,\\ Canberra, ACT 2601, Australia}

\author{K. G. H. Baldwin}
\affiliation{Laser Physics Centre, Research School of Physics, The Australian National University,\\ Canberra, ACT 2601, Australia}

\author{S. S. Hodgman}
\affiliation{Laser Physics Centre, Research School of Physics, The Australian National University,\\ Canberra, ACT 2601, Australia}

\author{A. G. Truscott}
\affiliation{Laser Physics Centre, Research School of Physics, The Australian National University,\\ Canberra, ACT 2601, Australia}

\date{\today}

\begin{abstract}

We present the detection of the highly forbidden \(\MetastableState \rightarrow \UpperState\) atomic transition in helium, the weakest transition observed in any neutral atom. Our measurements of the transition frequency, upper state lifetime, and transition strength agree well with published theoretical values, and can lead to tests of both QED contributions and different QED frameworks.
To measure such a weak transition, we developed two methods using ultracold metastable (\(\MetastableState\)) helium atoms: low background direct detection of excited then decayed atoms for sensitive measurement of the transition frequency and lifetime; and a pulsed atom laser heating measurement for determining the transition strength. These methods could possibly be applied to other atoms, providing new tools in the search for ultra-weak transitions and precision metrology.

\end{abstract}

\maketitle


The field of precision spectroscopy has made many foundational contributions to modern physics \cite{PhysRev.21.483,Landsberg1928,RAMAN1928,Lamb1947}, in particular through the development of quantum electrodynamics (QED) theory.
However, despite QED being one of the most rigorously tested theories in physics, there are still unknown factors and parameters, as shown by the recent ``proton radius puzzle" that required a re-assessment of the proton radius \cite{Pohl2010,Antognini417,Bezginov1007,Xiong2019}. This leads to an imperative to test QED at the highest precision using independent methods, in order to better understand its limitations.
Advances in laser technology have enabled the detection of an increasingly wide array of atomic transitions, including extremely weak atomic spectral lines from so-called forbidden transitions, which within a given approximation, e.g. the electric dipole approximation, strictly cannot occur. However, in reality such transitions do occur, but at extremely low rates.
The strength of an atomic transition is characterized by the Einstein \(A\) coefficient (the transition rate), which is challenging to either calculate or measure accurately. However, in some atomic systems the Einstein \(A\) coefficient has a significant and potentially measurable contribution from QED effects \cite{PhysRevA.79.032515}. Hence, measurements of the Einstein \(A\) coefficient can provide a test of QED, completely independent of, for example, the measurement of atomic energy intervals. Note that while there are other means of measuring transition rate information in atomic systems in order to test QED, such as the tune-out frequency (the frequency at which the atomic polarisability vanishes \cite{PhysRevA.75.053612,PhysRevLett.115.043004,PhysRevA.99.040502}), they often relate to the ratio of strong transition rates between multiple states. Thus these techniques do not measure the isolated strength of a single transition, which can provide additional insights and be important for specific applications, nor are they useful for measuring or constraining ultra weak transitions \cite{Pickering11}.

\begin{figure}[b]
    \centering
    \includegraphics[width=\linewidth]{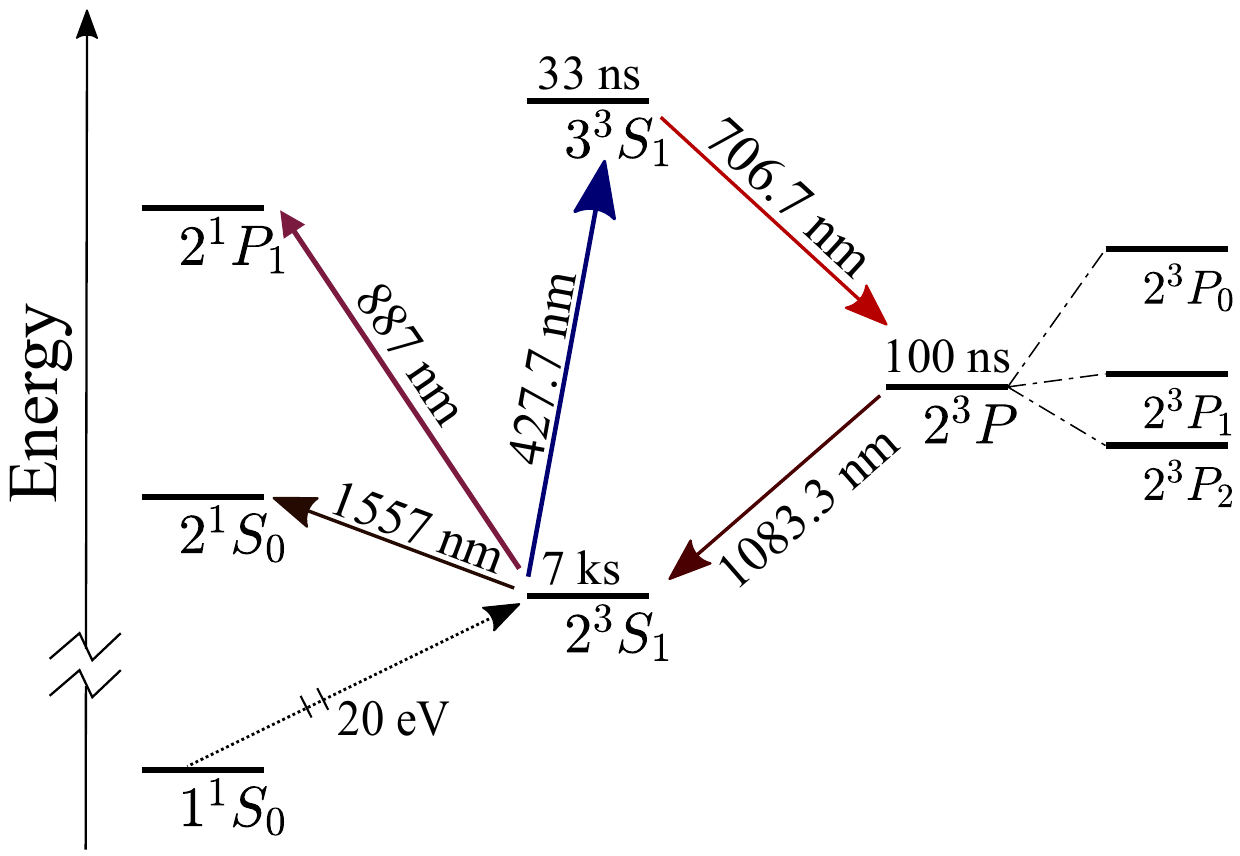}
    \caption{Partial atomic level scheme for helium. Level splittings are not to scale. The transition of interest, \(\MetastableState \rightarrow \UpperState\), is at \(427.7\)~nm (blue arrow), along with the dominant decay path from the \(\UpperState\) state (\(706.7\)~nm, red arrow). Relevant excited state lifetimes and transition wavelengths are also indicated.
    }
    \label{fig:level_diagram}
\end{figure}

A favoured test bed of QED models is the helium atom, where the two-electron structure is simple enough that theoretical calculations of many parameters can be determined to great precision. Decades of work on \(^3\)He and \(^4\)He systems have led to many advances, such as an improved measurement of the ground state Lamb shift \cite{PhysRevLett.105.063001,PhysRevLett.80.3475}, the fine structure constant \cite{PhysRevLett.105.123001,PhysRevLett.121.143002}, and both the alpha and helion particle charge radius \cite{Rengelink2018,PhysRevLett.108.143001}. There have also been a number of recent advancements specifically in precision spectroscopy of forbidden
transitions in the helium atom.
For instance the \(\MetastableState \rightarrow 2^{1\!}P_1\) transition (see Fig. \ref{fig:level_diagram}), which is forbidden as it violates spin conservation and has a predicted Einstein \(A\) value of $A=1.4432$~$\text{s}^{\text{-}1}$ \cite{Drake_2007}, was first observed by Notermans \textit{et al.} to a precision of \(0.5\)~MHz \cite{PhysRevLett.112.253002}. 

Furthermore, a second extremely weak helium transition of interest is the singlet to triplet ground state transition of metastable helium (He\(^*\)) \(\MetastableState \rightarrow \WimState\)  (see Fig. \ref{fig:level_diagram}), which is doubly forbidden, as it links a triplet to a singlet state, and \(\Delta l = 0\). This transition has a predicted Einstein \(A\) coefficient ranging from $A=6.1\times10^{\text{-}8}$~$\text{s}^{\text{-}1}$ \cite{ISI:000071951300016} to $A=1.5\times10^{\text{-}7}$~ $\text{s}^{\text{-}1}$ \cite{PhysRevA.15.154}, but the transition rate is yet to be measured. 
An experimental measurement of the transition frequency was carried out by van Rooij \textit{et al.} to a precision of \(2\)~kHz for both $^3$He and $^4$He \cite{vanRooij196}. Subsequent measurements by Rengelink \textit{et al.} improved the precision to \(0.2\)~kHz by using a magic wavelength trap \cite{Rengelink2018}, providing a new test of QED and nuclear structure calculations, including a determination of the nuclear charge radius. Of further note are the frequency measurements of seven of the transitions between the \(2^{3\!}S\) and \(2^{3\!}P\) hyperfine manifolds in \(^3\)He by Cancio Pastor \textit{et al.} to the order of \(1\)~kHz \cite{PhysRevLett.108.143001}. This provided a value of the difference of the squared nuclear charge radii of \(^3\)He and $^4$He, which differed by \(4\sigma\) from that derived by van Rooij \textit{et al.}, exemplifying the need to perform different types of experiments to properly constrain QED theory.

Another transition in helium that until now has not been detected experimentally is the strongly forbidden \(\MetastableState \rightarrow \UpperState\) transition (see Fig. \ref{fig:level_diagram}), for which \(\Delta l = 0\), and it is hence electric dipole forbidden. It is excited via the magnetic dipole interaction using light with a predicted wavelength of \(\sim\)\(427.7\)~nm \cite{Drake2006}. 
There are unresolved conflicting theoretical predictions for the Einstein \(A\) coefficient of this transition. Derevianko \textit{et al.} predict \(A=1.17\times10^{\text{-}8}\)~\(\text{s}^{\text{-}1}\) \cite{PhysRevA.58.4453}, while a calculation by \L{}ach \textit{et al.} gives \(A=6.48\times10^{\text{-}9}\)~\(\text{s}^{\text{-}1}\) \cite{PhysRevA.64.042510}, which states in reference to the differing values \textit{``This discrepancy does not have experimental impact since this rate is too small \ldots to be measured"} \cite{PhysRevA.64.042510}. An accurate measurement of the Einstein \(A\) coefficient for this transition would provide insight into the validity and limitations of the different approaches within QED theory. These calculations also indicate that this transition rate would be the weakest ever measured in a neutral atom, and only slightly stronger than the weakest measured transition rate in an ion: the electric-octupole transition in \(^{172}\)Yb+, which is the longest lived at \(8.4\)~years, \textit{i.e.} \(A = 3.8\times10^{\text{-}9}\)~\(\text{s}^{\text{-}1}\) (theory \cite{PhysRevLett.81.3345}), or \(10^{+7}_{-4}\)~years, equivalently \(A = 3^{+2}_{-1}\times10^{\text{-}9}\)~\(\text{s}^{\text{-}1}\) (experiment \cite{PhysRevLett.78.1876}).


In this work we present the first detection of the \(\MetastableState \rightarrow \UpperState\) transition in $^4$He. We develop two novel techniques for the measurement of ultra-weak transitions and use them to determine the transition frequency,
Einstein \(A\) coefficient and excited state lifetime. The first method uses a Bose-Einstein condensate (BEC) and directly detects atoms which absorb a photon and escape a shallow trap. While this method is highly sensitive and is ideal for the determination of the transition frequency and linewidth, the uncertainty in the collection efficiency necessitates an independent approach for determining the Einstein \(A\) coefficient. To this end we developed a second method, which measures the heating rate of a trapped thermal cloud due to the absorption and subsequent re-emission of photons from a probe beam. From this the Einstein \(A\) coefficient can be extracted. While the use of heating due to photon recoil to detect an excitation has been used for great precision and sensitivity in ion spectroscopy \cite{PhysRevLett.78.1876,Wan2014,PhysRevLett.115.053003,Guggemos_2019} this is the first time such a technique has been utilised in a neutral atom system.
Similar technique could possibly be used to search for other weak transitions which have applications in astronomy and state-of-the-art technologies, such as atomic clocks \cite{detection_note}.

\begin{figure}[t]
\includegraphics[width=\linewidth]{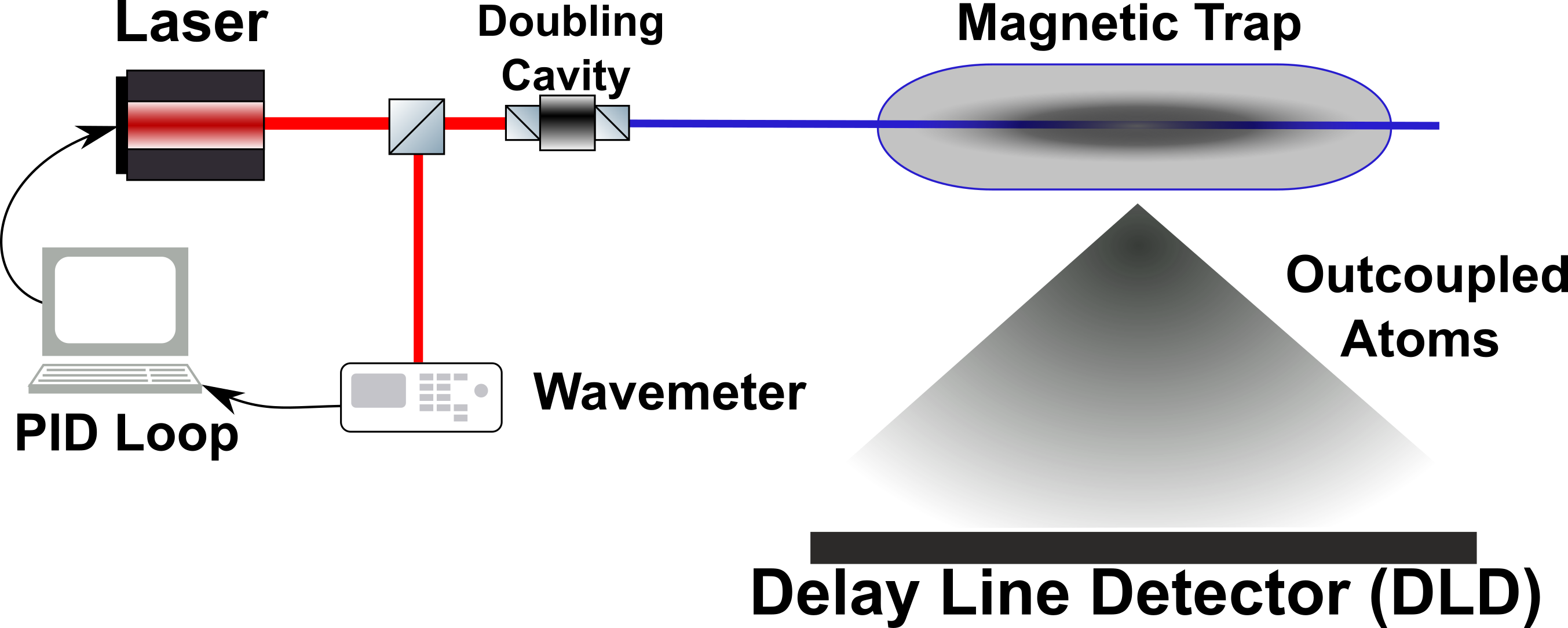}
\caption{\label{fig:exp_setup}Diagram of the experimental setup. A BEC is produced and then held in a magnetic trap. The laser light is focused onto the atoms in the trap and when an atom absorbs one of the photons it will most likely leave the trap, with some high probability of it landing on the detector.}
\end{figure}

To measure the transition frequency and linewidth we start with a BEC of \(\sim\)\(10^6\) He\(^*\) atoms trapped in the long-lived \(\MetastableState\) excited state \cite{PhysRevLett.103.053002}, prepared via a combination of laser and evaporative cooling in a magnetic bi-planar quadrupole Ioffe trap \cite{Dall2007}. The atoms are prepared in the \(m_J=+1\) magnetic substate, as this is the only magnetically trapped state \cite{Dall2007}. The atoms are detected after falling onto an 80 mm diameter micro-channel plate and delay line detector (DLD) \cite{Manning:10} (see remark \cite{detection_note} for extensions to other atoms), located approximately 850 mm below the trap centre (Fig.~\ref{fig:exp_setup}). 

To address the \(\MetastableState \rightarrow \UpperState\) transition we illuminate the atoms with a probe beam from a laser and doubling cavity that is tuneable around 427.7nm \cite{SOMs}.  The frequency of the laser was stabilised using a feedback loop to a wavemeter with \(2\) MHz absolute accuracy, which was periodically calibrated to a known cesium crossover transition (see \cite{SOMs} for further detail). After passing through an optical fibre, the probe beam is focused and aligned along the weak axis of the trap, see Fig.~\ref{fig:exp_setup} for diagram of experimental setup. We then perform differential measurements between the laser applied and a reference shot with the laser blocked.

The transition is detected by directly measuring small numbers, on the order of \(10^2\), of atoms that absorb the probe laser light during a \(25\)~s exposure time. 
When the wavelength of a \(\sigma^-\) polarised probe laser beam is resonant with the \(\MetastableState \rightarrow \UpperState\) transition, the \(427.7\)~nm photon excites the atom from the \(\MetastableState , m_J=+1\) state to the \(\UpperState , m_J=0\) state and the atom receives a momentum recoil. 
The vast majority of these excited atoms then decay within $\sim$30~ns, emitting a photon at \(706.7\)~nm to one of the \(\MidState\) states, then within $\sim$100~ns decay via the \(1083\)~nm transitions to the \(\MetastableState\) state (see Fig.~\ref{fig:level_diagram}). This is because all other transitions from \(\UpperState\) and the \(\MidState\) states are forbidden: hence less than 1 in \(10^4\) atoms will decay to non-\(\MetastableState\) states \cite{PhysRevLett.100.023001,Hodgman2009a}.

Atoms will hence on average end up distributed among the magnetic sublevels \(m_J=(-1,0, 1)\) of the \(\MetastableState\) state with a fractional population of (24\%,52\%,24\%) based on the relevant transition Clebsch-Gordan coefficients. 
The 76\% of atoms that decay to the untrapped \(\MetastableState, \, m_J=0\, \text{or} \, -1\) states leave the trap immediately and fall under the influence of gravity onto the DLD, with the chance that they will collide with other atoms while leaving the BEC \cite{SOMs}. The remaining \(24\%\) of excited atoms that have decayed to the \(\MetastableState, m_J=+1\) state will be re-trapped.
\begin{figure}[b]
\centering
\includegraphics[width=\linewidth]{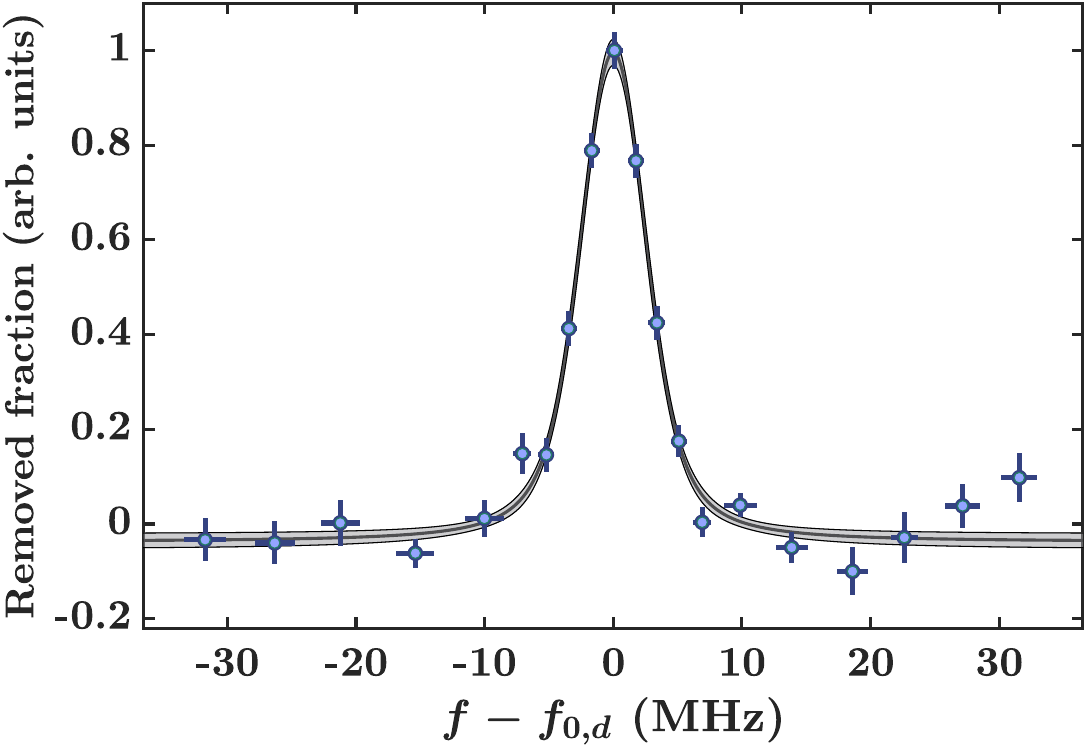}
\caption{(Color online) The normalised excited fraction as a function of applied laser frequency (relative to the fitted centered frequency \(f_{0,d} = 700,939,271.64(8)\)~MHz where quoted error is purely statistical). Vertical and horizontal bars indicate the uncertainty in their respective axis \cite{standard_error_note}. Data has been binned for viewing, where the width of the bin used to calculate each point is varied to compensate for the varying density of sample points. The black line is a Voigt fit to the data, with the grey shaded region indicating the confidence interval.  The parameters of the fit are \(\sigma=1.9(4)\)~MHz (standard deviation of the Gaussian component) and \(\gamma=3.2(10)\)~MHz (scale parameter of the Lorentzian component), corresponding to an excited state lifetime of \(50(20)\)~ns. The peak signal represents an excited fraction of 0.34\% for a total energy of applied photons of \(0.65\)~J.
}
\label{fig:427nm_signal}
\end{figure}

After the probe beam is switched off, the remaining atoms in the magnetic trap are outcoupled with pulses of broadband RF radiation.
This transfers all atoms from the trap into a coherent beam of atoms, known as an atom laser \cite{Manning:10,Henson2018}, allowing the total number of remaining atoms in the trap to be measured, while avoiding detector saturation. The ratio of excited and lost atoms to remaining atoms can hence be determined, which is less sensitive to total BEC number fluctuations from shot-to-shot.


For each laser wavelength, \(\sim\)\(215\) shots are taken with the probe beam applied and \(\sim\)\(50\) with it blocked as a reference, from which the normalised excitation probability per photon per unit time \cite{SOMs} is extracted. The excited fractions for a range of frequencies around the transition are shown in Fig.~\ref{fig:427nm_signal}.  At resonance, we measure a peak signal corresponding to 0.34\% of the total atoms excited per \(\sim\)\(10^{18}\) applied photons (for details on the beam shape and power in relation to the atom sample see \cite{SOMs}).  Note that the signal in Fig.~\ref{fig:427nm_signal} decays to a negative value far from the transition.  We speculate that this is due to the off-resonant repulsive dipole potential of the probe beam on the atoms, which causes a deflection of atoms such that they miss the detector, compared to the reference case.  While this effect is measurable, it has a negligible effect on the line shape compared to the other sources of error \cite{SOMs}. 



The centre of the corresponding Lorentzian fit gives a measured transition frequency of \(f_{0,d}=700,939,271.64(8)\)~MHz, with subscript \(d\) referring to the direct detection method and with only the statistical uncertainty shown.  After applying relevant systematic corrections (as listed with the full error budget in Tab.~\ref{tab:corrections}), this yields a final value of \(f_{0,d}^{shifted}=700,939,271(5)\)~MHz.  This agrees very well with the most recent published value in the literature of \(700,939,269(8)\)~MHz \cite{Drake2006}, with our uncertainty smaller than that of theory.  The Lorentzian width of the peak, derived from the Voigt fit (see Fig.~\ref{fig:427nm_signal}), also allows the state lifetime of the $3^{3\!}S_1$ state to be determined.  We estimate an excited state lifetime of \(\tau = 50(20)\)~ns, which compares well to the theoretical value of \(35.9(2)\)~ns \cite{SOMs}. We also find that the sensitivty of this method is such that an Einstein \(A\) value of \(\approx 7 \times10^{-11}\)~\(\text{s}^{\text{-}1}\) could be observed with a SNR of unity given one day of interrogation \cite{SOMs}.

\begin{table}[b]
    \centering
    \begin{tabular}{@{\extracolsep{\fill}} l @{\extracolsep{\fill}}>{\centering}p{0.09\textwidth} @{\extracolsep{\fill}}>{\centering}p{0.09\textwidth} @{\extracolsep{\fill}}>{\centering}p{0.07\textwidth} @{\extracolsep{\fill}}>{\centering\arraybackslash}p{0.07\textwidth}}
    \toprule
    \toprule
        Value & \multicolumn{2}{c}{Systematic} & \multicolumn{2}{c}{Unc (MHz)}\\
        &\multicolumn{2}{c}{Freq Shift (MHz)}&\multicolumn{2}{c}{}\\
        \hline
         &\(f_{0,d}\)&\(f_{0,h}\)&\(f_{0,d}\)&\(f_{0,h}\)\\
        \hline
         Zeeman shift &\multicolumn{2}{c}{ \(-1.715\) } & \multicolumn{2}{c}{0.003} \\
         AC Stark shift & 6.9 &5.9  & 1.5 & 1.6\\
         DC Stark shift & \multicolumn{2}{c}{\(<10^{-6}\)} & \multicolumn{2}{c}{-}\\
         Mean field shift & \multicolumn{2}{c}{\(<0.01\)} & \multicolumn{2}{c}{-}\\
         Recoil shift & \multicolumn{2}{c}{0.273}  & \multicolumn{2}{c}{\(<\)0.001}  \\
         Cesium Cell offset & \multicolumn{2}{c}{} & \multicolumn{2}{c}{}\\
         \, \, - AC Stark shift & \multicolumn{2}{c}{\(-1.9\)} & \multicolumn{2}{c}{0.4} \\
         \, \, - Pressure shift & \multicolumn{2}{c}{\(<0.006\)} & \multicolumn{2}{c}{-} \\
         Wavemeter & \multicolumn{2}{c}{\(-3.0\)} & \multicolumn{2}{c}{4.1}\\
         Statistical & \multicolumn{2}{c}{-}  & 0.08 & 0.6\\
         \hline
         Total & 0.6 & -0.4 & 4.4 & 4.5\\
    \bottomrule
    \bottomrule
    \end{tabular}
    \caption{Systematic shifts, corrections, and uncertainties to measured frequency values from the direct detection method \(f_{0,d}\) and the heating method \(f_{0,h}\). Note that uncertainties are added in quadrature.}
    \label{tab:corrections}
\end{table}


To measure the transition strength, we employ a different experimental technique that determines the heating of the cloud induced by the photon recoil of absorbed and emitted photons from the probe beam. The thermal cloud has an initial temperature of order \(1\)~\(\mu\)K.  We use a minimally-destructive spectrally broad RF pulse to remove \(\sim\)2\% of the atoms from the trap. The pulses are approximately 20~\(\mu\)s in length, and hence have a Fourier width of \(\sim 300\)~kHz \cite{Manning:10}, which ensures uniform outcoupling throughout the trap. The time-of-flight profile recorded on the DLD in the far field will represent the momentum profile of the trapped atoms \cite{Yavin2002}. As the temperature of the atoms is significantly above the condensation temperature, \(T_c \sim 150\)~nK, the temperature was found by fitting each profile with a Boltzmann distribution \cite{SOMs}.
%
\begin{figure}[t]
    \centering
    \includegraphics[width=\linewidth]{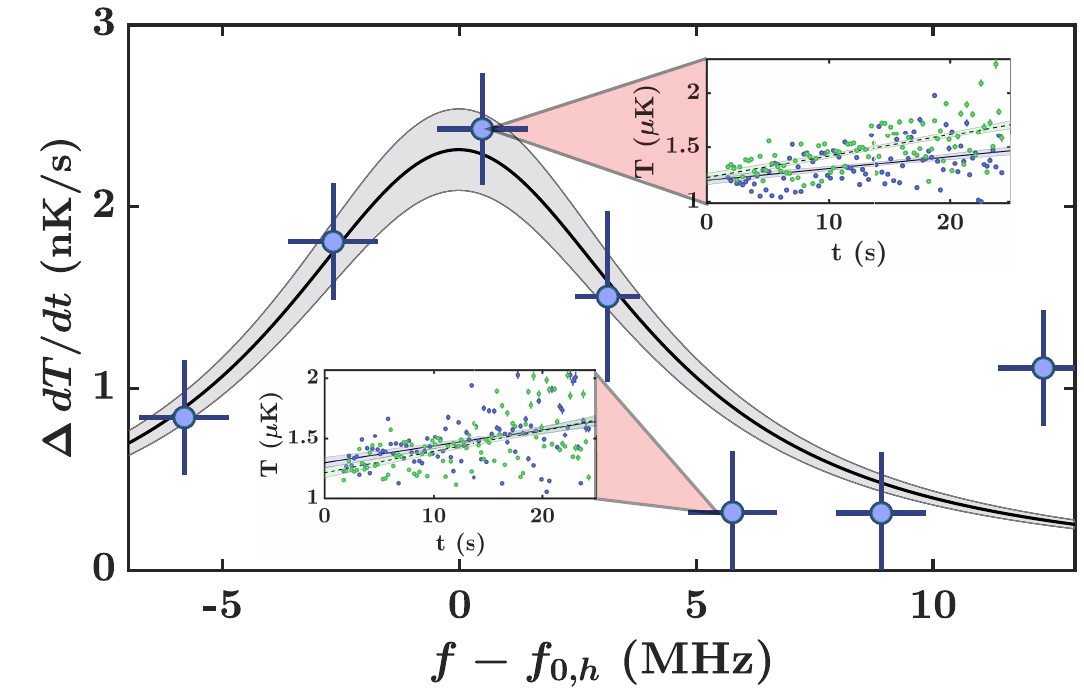} 
    \caption{(Color online) Increase in heating rate as a function of applied laser frequency, relative to fitted frequency center \(f_{0,h} = 700,939,270.9(6)\)~MHz, with quoted error purely statistical. Data has been binned in frequency for clarity, with vertical and horizontal bars indicating uncertainty in the respective axis \cite{standard_error_note}. The solid black line represents a Voigt fit with \(\sigma=1.6(9)\)~MHz, and \(\gamma=4(3)\)~MHz. The insets show a comparison of heating rates at the respective frequencies, with the dashed (green) line indicating a run with the laser light applied and the solid (blue) line indicating a reference run.
    }
    \label{fig:427nm_signal_heating}
\end{figure}
By repeatedly outcoupling small numbers of atoms (the full sequence uses 95 pulses each spaced 240~ms apart), the temperature of the trapped thermal cloud can be estimated as a function of time, and thus a heating rate determined. Comparison of the measured heating rate when the probe beam is present to when it is blocked allows an estimate of the heating rate due to the probe beam.  The difference in the heating rates between probe and reference is shown in Fig.~\ref{fig:427nm_signal_heating} as a function of laser frequency, which gives a fitted peak frequency for this method of \(f_{0,h}^{shifted}=700,939,270.9(6)\)~MHz, (with subscript \(h\) referring to the heating method, and with the statistical uncertainty shown). After applying appropriate systematic frequency shifts (see Tab.~\ref{tab:corrections} \cite{SOMs}), the final value for the transition frequency is \(f_{0,h} = 700,939,271(5)\)~MHz, and the excited state lifetime is \(40(30)\)~ns. Both agree within uncertainty with the values measured by the direct detection method.

We calculate the Einstein \(A\) coefficient from the measured heating rate and the heat capacity of a harmonically trapped Bose gas \cite{SOMs}. The resultant value is \(A =7(4)\times 10^{\text{-}9}\)~\(\text{s}^{\text{-}1}\), compared to the most recent theoretical value of \(A=6.48\times 10^{\text{-}9}\)~\(\text{s}^{\text{-}1}\) \cite{PhysRevA.64.042510}.

\renewcommand\arraystretch{1.0}
\begin{table}[b]
    \centering
    \begin{tabular}{@{\extracolsep{\fill}} l @{\extracolsep{\fill}} c @{\extracolsep{\fill}} c @{\extracolsep{\fill}} c}
    \toprule
    \toprule
         \multicolumn{1}{l}{Method} & \multicolumn{1}{c}{Center}  & \multicolumn{1}{c}{\(\UpperState\) State}  & \multicolumn{1}{c}{Einstein \(A\)}  \\
          & \multicolumn{1}{c}{Freq (MHz)} & \multicolumn{1}{c}{Lifetime (ns)} & \multicolumn{1}{c}{Coeff (\(10^{\text{-}9}\)s\(^{\text{-}1}\))}\\
         \hline 
         Direct & \(700,939,271(5)\) & \(50(20)\) & - \\
         Heating & \(700,939,271(5)\) & \(40(30)\) & \(7(4)\) \\
         Theory & \(700,939,269(8)\)\cite{Drake2006} & \(35.9(2)\)\cite{SOMs} &\(6.48\)\cite{PhysRevA.64.042510}, \(11.7\)\cite{PhysRevA.58.4453}\\
    \bottomrule
    \bottomrule
    \end{tabular}
    \caption{Summary table of experimentally measured values, including all systematic corrections, for the \(\MetastableState \rightarrow \UpperState\) transition in Helium, with the most recent theoretical calculations for comparison.}
    \label{tab:summary}
\end{table}

Our results for the transition frequency using both methods compare well with the most recent theoretical value in the literature (see Tab.~\ref{tab:summary}), and our experimental uncertainty is comparable to that of the current QED theory calculation.  A further consequence of our measurement of the \(\MetastableState \rightarrow \UpperState\) transition wavelength is that it constrains the \(2^{3\!}P_1 \rightarrow \UpperState\) transition frequency to be \(424,202,774(5)\)~MHz, using the extremely accurately measured \(\MetastableState \rightarrow 2^{3\!}P_1\) transition frequency \cite{PhysRevLett.119.263002}. Further, the experimental Einstein \(A\) coefficient also agrees within error with both of the most recent theoretical published values \cite{PhysRevA.58.4453,PhysRevA.64.042510}, although it is not sufficiently sensitive to resolve the difference between them. Nonetheless, the measurement of transition strengths is important as an alternative test for QED, as there are few techniques which can be compared to energy level measurements, and thus further investigation is warranted.

In conclusion, we have demonstrated a novel sensitive method for measuring and characterising spectroscopic transitions in helium that could in principle be extended to other metastable atoms, particularly those that are used in ultracold gas experiments \cite{RevModPhys.84.175}. This has allowed us to detect the weakest transition ever observed in a neutral atom.  The techniques are based upon momentum detection of atoms, separating them from most other techniques in the literature which are usually based upon measuring change in irradiance.  While our method agrees within experimental uncertainty with theory, by increasing the accuracy of the laser wavelength measurement  (e.g. via incorporating a frequency comb), we could reach a level of accuracy of \(<1\)~MHz, which would provide a challenge to improve state-of-the-art theoretical predictions.  Furthermore, by conducting similar measurements on $^3$He, isotope shifts could also be compared as a further test of QED predictions.


\begin{acknowledgments}
The authors would like to thank G. W. F. Drake, G. \L{}ach and K. Pachucki for helpful discussions, and C. J. Vale and S. Hoinka for the loan of the laser.
This work was supported through Australian Research
Council (ARC) Discovery Project Grants DP160102337 and DP180101093, 
as well as Linkage Project LE180100142.
K.F.T. and D.K.S. are supported by Australian Government Research Training Program (RTP) Scholarships.
S. S. H. is supported by ARC Discovery Early Career Researcher Award No. DE150100315.
\end{acknowledgments}

\bibliography{forbidden427}

\begin{thebibliography}{69}%
\makeatletter
\providecommand \@ifxundefined [1]{%
 \@ifx{#1\undefined}
}%
\providecommand \@ifnum [1]{%
 \ifnum #1\expandafter \@firstoftwo
 \else \expandafter \@secondoftwo
 \fi
}%
\providecommand \@ifx [1]{%
 \ifx #1\expandafter \@firstoftwo
 \else \expandafter \@secondoftwo
 \fi
}%
\providecommand \natexlab [1]{#1}%
\providecommand \enquote  [1]{``#1''}%
\providecommand \bibnamefont  [1]{#1}%
\providecommand \bibfnamefont [1]{#1}%
\providecommand \citenamefont [1]{#1}%
\providecommand \href@noop [0]{\@secondoftwo}%
\providecommand \href [0]{\begingroup \@sanitize@url \@href}%
\providecommand \@href[1]{\@@startlink{#1}\@@href}%
\providecommand \@@href[1]{\endgroup#1\@@endlink}%
\providecommand \@sanitize@url [0]{\catcode `\\12\catcode `\$12\catcode
  `\&12\catcode `\#12\catcode `\^12\catcode `\_12\catcode `\%12\relax}%
\providecommand \@@startlink[1]{}%
\providecommand \@@endlink[0]{}%
\providecommand \url  [0]{\begingroup\@sanitize@url \@url }%
\providecommand \@url [1]{\endgroup\@href {#1}{\urlprefix }}%
\providecommand \urlprefix  [0]{URL }%
\providecommand \Eprint [0]{\href }%
\providecommand \doibase [0]{https://doi.org/}%
\providecommand \selectlanguage [0]{\@gobble}%
\providecommand \bibinfo  [0]{\@secondoftwo}%
\providecommand \bibfield  [0]{\@secondoftwo}%
\providecommand \translation [1]{[#1]}%
\providecommand \BibitemOpen [0]{}%
\providecommand \bibitemStop [0]{}%
\providecommand \bibitemNoStop [0]{.\EOS\space}%
\providecommand \EOS [0]{\spacefactor3000\relax}%
\providecommand \BibitemShut  [1]{\csname bibitem#1\endcsname}%
\let\auto@bib@innerbib\@empty
\bibitem [{\citenamefont {Compton}(1923)}]{PhysRev.21.483}%
  \BibitemOpen
  \bibfield  {author} {\bibinfo {author} {\bibfnamefont {A.~H.}\ \bibnamefont
  {Compton}},\ }\bibfield  {title} {\bibinfo {title} {A quantum theory of the
  scattering of {X}-rays by light elements},\ }\href
  {https://doi.org/10.1103/PhysRev.21.483} {\bibfield  {journal} {\bibinfo
  {journal} {Phys. Rev.}\ }\textbf {\bibinfo {volume} {21}},\ \bibinfo {pages}
  {483} (\bibinfo {year} {1923})}\BibitemShut {NoStop}%
\bibitem [{\citenamefont {Landsberg}\ and\ \citenamefont
  {Mandelstam}(1928)}]{Landsberg1928}%
  \BibitemOpen
  \bibfield  {author} {\bibinfo {author} {\bibfnamefont {G.}~\bibnamefont
  {Landsberg}}\ and\ \bibinfo {author} {\bibfnamefont {L.}~\bibnamefont
  {Mandelstam}},\ }\bibfield  {title} {\bibinfo {title} {A novel effect of
  light scattering in crystals},\ }\href {https://doi.org/10.1007/BF01506807}
  {\bibfield  {journal} {\bibinfo  {journal} {Naturwissenschaften}\ }\textbf
  {\bibinfo {volume} {16}},\ \bibinfo {pages} {557} (\bibinfo {year}
  {1928})}\BibitemShut {NoStop}%
\bibitem [{\citenamefont {Raman}\ and\ \citenamefont
  {Krishnan}(1928)}]{RAMAN1928}%
  \BibitemOpen
  \bibfield  {author} {\bibinfo {author} {\bibfnamefont {C.~V.}\ \bibnamefont
  {Raman}}\ and\ \bibinfo {author} {\bibfnamefont {K.~S.}\ \bibnamefont
  {Krishnan}},\ }\bibfield  {title} {\bibinfo {title} {The optical analogue of
  the {C}ompton effect},\ }\href {https://doi.org/10.1038/121711a0} {\bibfield
  {journal} {\bibinfo  {journal} {Nature}\ }\textbf {\bibinfo {volume} {121}},\
  \bibinfo {pages} {711} (\bibinfo {year} {1928})}\BibitemShut {NoStop}%
\bibitem [{\citenamefont {Lamb}\ and\ \citenamefont
  {Retherford}(1947)}]{Lamb1947}%
  \BibitemOpen
  \bibfield  {author} {\bibinfo {author} {\bibfnamefont {W.~E.}\ \bibnamefont
  {Lamb}}\ and\ \bibinfo {author} {\bibfnamefont {R.~C.}\ \bibnamefont
  {Retherford}},\ }\bibfield  {title} {\bibinfo {title} {Fine structure of the
  hydrogen atom by a microwave method},\ }\href
  {https://doi.org/10.1103/PhysRev.72.241} {\bibfield  {journal} {\bibinfo
  {journal} {Phys. Rev.}\ }\textbf {\bibinfo {volume} {72}},\ \bibinfo {pages}
  {241} (\bibinfo {year} {1947})}\BibitemShut {NoStop}%
\bibitem [{\citenamefont {Pohl}\ \emph {et~al.}(2010)\citenamefont {Pohl},
  \citenamefont {Antognini}, \citenamefont {Nez}, \citenamefont {Amaro} \emph
  {et~al.}}]{Pohl2010}%
  \BibitemOpen
  \bibfield  {author} {\bibinfo {author} {\bibfnamefont {R.}~\bibnamefont
  {Pohl}}, \bibinfo {author} {\bibfnamefont {A.}~\bibnamefont {Antognini}},
  \bibinfo {author} {\bibfnamefont {F.}~\bibnamefont {Nez}}, \bibinfo {author}
  {\bibfnamefont {F.~D.}\ \bibnamefont {Amaro}}, \emph {et~al.},\ }\bibfield
  {title} {\bibinfo {title} {The size of the proton},\ }\href
  {https://doi.org/10.1038/nature09250} {\bibfield  {journal} {\bibinfo
  {journal} {Nature}\ }\textbf {\bibinfo {volume} {466}},\ \bibinfo {pages}
  {213} (\bibinfo {year} {2010})}\BibitemShut {NoStop}%
\bibitem [{\citenamefont {Antognini}\ \emph {et~al.}(2013)\citenamefont
  {Antognini}, \citenamefont {Nez}, \citenamefont {Schuhmann}, \citenamefont
  {Amaro} \emph {et~al.}}]{Antognini417}%
  \BibitemOpen
  \bibfield  {author} {\bibinfo {author} {\bibfnamefont {A.}~\bibnamefont
  {Antognini}}, \bibinfo {author} {\bibfnamefont {F.}~\bibnamefont {Nez}},
  \bibinfo {author} {\bibfnamefont {K.}~\bibnamefont {Schuhmann}}, \bibinfo
  {author} {\bibfnamefont {F.~D.}\ \bibnamefont {Amaro}}, \emph {et~al.},\
  }\bibfield  {title} {\bibinfo {title} {Proton structure from the measurement
  of 2{S}-2{P} transition frequencies of muonic hydrogen},\ }\href
  {https://doi.org/10.1126/science.1230016} {\bibfield  {journal} {\bibinfo
  {journal} {Science}\ }\textbf {\bibinfo {volume} {339}},\ \bibinfo {pages}
  {417} (\bibinfo {year} {2013})}\BibitemShut {NoStop}%
\bibitem [{\citenamefont {Bezginov}\ \emph {et~al.}(2019)\citenamefont
  {Bezginov}, \citenamefont {Valdez}, \citenamefont {Horbatsch}, \citenamefont
  {Marsman}, \citenamefont {Vutha},\ and\ \citenamefont
  {Hessels}}]{Bezginov1007}%
  \BibitemOpen
  \bibfield  {author} {\bibinfo {author} {\bibfnamefont {N.}~\bibnamefont
  {Bezginov}}, \bibinfo {author} {\bibfnamefont {T.}~\bibnamefont {Valdez}},
  \bibinfo {author} {\bibfnamefont {M.}~\bibnamefont {Horbatsch}}, \bibinfo
  {author} {\bibfnamefont {A.}~\bibnamefont {Marsman}}, \bibinfo {author}
  {\bibfnamefont {A.~C.}\ \bibnamefont {Vutha}},\ and\ \bibinfo {author}
  {\bibfnamefont {E.~A.}\ \bibnamefont {Hessels}},\ }\bibfield  {title}
  {\bibinfo {title} {A measurement of the atomic hydrogen lamb shift and the
  proton charge radius},\ }\href {https://doi.org/10.1126/science.aau7807}
  {\bibfield  {journal} {\bibinfo  {journal} {Science}\ }\textbf {\bibinfo
  {volume} {365}},\ \bibinfo {pages} {1007} (\bibinfo {year}
  {2019})}\BibitemShut {NoStop}%
\bibitem [{\citenamefont {Xiong}\ \emph {et~al.}(2019)\citenamefont {Xiong},
  \citenamefont {Gasparian}, \citenamefont {Gao} \emph {et~al.}}]{Xiong2019}%
  \BibitemOpen
  \bibfield  {author} {\bibinfo {author} {\bibfnamefont {W.}~\bibnamefont
  {Xiong}}, \bibinfo {author} {\bibfnamefont {A.}~\bibnamefont {Gasparian}},
  \bibinfo {author} {\bibfnamefont {H.}~\bibnamefont {Gao}}, \emph {et~al.},\
  }\bibfield  {title} {\bibinfo {title} {A small proton charge radius from
  an~electron{\textendash}proton scattering experiment},\ }\href
  {https://doi.org/10.1038/s41586-019-1721-2} {\bibfield  {journal} {\bibinfo
  {journal} {Nature}\ }\textbf {\bibinfo {volume} {575}},\ \bibinfo {pages}
  {147} (\bibinfo {year} {2019})}\BibitemShut {NoStop}%
\bibitem [{\citenamefont {Andreev}\ \emph {et~al.}(2009)\citenamefont
  {Andreev}, \citenamefont {Labzowsky},\ and\ \citenamefont
  {Plunien}}]{PhysRevA.79.032515}%
  \BibitemOpen
  \bibfield  {author} {\bibinfo {author} {\bibfnamefont {O.~Y.}\ \bibnamefont
  {Andreev}}, \bibinfo {author} {\bibfnamefont {L.~N.}\ \bibnamefont
  {Labzowsky}},\ and\ \bibinfo {author} {\bibfnamefont {G.}~\bibnamefont
  {Plunien}},\ }\bibfield  {title} {\bibinfo {title} {{Q}{E}{D} calculation of
  transition probabilities in two-electron ions},\ }\href
  {https://doi.org/10.1103/PhysRevA.79.032515} {\bibfield  {journal} {\bibinfo
  {journal} {Phys. Rev. A}\ }\textbf {\bibinfo {volume} {79}},\ \bibinfo
  {pages} {032515} (\bibinfo {year} {2009})}\BibitemShut {NoStop}%
\bibitem [{\citenamefont {LeBlanc}\ and\ \citenamefont
  {Thywissen}(2007)}]{PhysRevA.75.053612}%
  \BibitemOpen
  \bibfield  {author} {\bibinfo {author} {\bibfnamefont {L.~J.}\ \bibnamefont
  {LeBlanc}}\ and\ \bibinfo {author} {\bibfnamefont {J.~H.}\ \bibnamefont
  {Thywissen}},\ }\bibfield  {title} {\bibinfo {title} {Species-specific
  optical lattices},\ }\href {https://doi.org/10.1103/PhysRevA.75.053612}
  {\bibfield  {journal} {\bibinfo  {journal} {Phys. Rev. A}\ }\textbf {\bibinfo
  {volume} {75}},\ \bibinfo {pages} {053612} (\bibinfo {year}
  {2007})}\BibitemShut {NoStop}%
\bibitem [{\citenamefont {Henson}\ \emph {et~al.}(2015)\citenamefont {Henson},
  \citenamefont {Khakimov}, \citenamefont {Dall}, \citenamefont {Baldwin},
  \citenamefont {Tang},\ and\ \citenamefont
  {Truscott}}]{PhysRevLett.115.043004}%
  \BibitemOpen
  \bibfield  {author} {\bibinfo {author} {\bibfnamefont {B.~M.}\ \bibnamefont
  {Henson}}, \bibinfo {author} {\bibfnamefont {R.~I.}\ \bibnamefont
  {Khakimov}}, \bibinfo {author} {\bibfnamefont {R.~G.}\ \bibnamefont {Dall}},
  \bibinfo {author} {\bibfnamefont {K.~G.~H.}\ \bibnamefont {Baldwin}},
  \bibinfo {author} {\bibfnamefont {L.-Y.}\ \bibnamefont {Tang}},\ and\
  \bibinfo {author} {\bibfnamefont {A.~G.}\ \bibnamefont {Truscott}},\
  }\bibfield  {title} {\bibinfo {title} {Precision measurement for metastable
  helium atoms of the 413 nm tune-out wavelength at which the atomic
  polarizability vanishes},\ }\href
  {https://doi.org/10.1103/PhysRevLett.115.043004} {\bibfield  {journal}
  {\bibinfo  {journal} {Phys. Rev. Lett.}\ }\textbf {\bibinfo {volume} {115}},\
  \bibinfo {pages} {043004} (\bibinfo {year} {2015})}\BibitemShut {NoStop}%
\bibitem [{\citenamefont {Zhang}\ \emph {et~al.}(2019)\citenamefont {Zhang},
  \citenamefont {Wu}, \citenamefont {Zhang}, \citenamefont {Tang},
  \citenamefont {Zhang}, \citenamefont {Baldwin},\ and\ \citenamefont
  {Shi}}]{PhysRevA.99.040502}%
  \BibitemOpen
  \bibfield  {author} {\bibinfo {author} {\bibfnamefont {Y.-H.}\ \bibnamefont
  {Zhang}}, \bibinfo {author} {\bibfnamefont {F.-F.}\ \bibnamefont {Wu}},
  \bibinfo {author} {\bibfnamefont {P.-P.}\ \bibnamefont {Zhang}}, \bibinfo
  {author} {\bibfnamefont {L.-Y.}\ \bibnamefont {Tang}}, \bibinfo {author}
  {\bibfnamefont {J.-Y.}\ \bibnamefont {Zhang}}, \bibinfo {author}
  {\bibfnamefont {K.~G.~H.}\ \bibnamefont {Baldwin}},\ and\ \bibinfo {author}
  {\bibfnamefont {T.-Y.}\ \bibnamefont {Shi}},\ }\bibfield  {title} {\bibinfo
  {title} {{Q}{E}{D} and relativistic nuclear recoil corrections to the 413-nm
  tune-out wavelength for the 2 $^{3}{S}_{1}$ state of helium},\ }\href
  {https://doi.org/10.1103/PhysRevA.99.040502} {\bibfield  {journal} {\bibinfo
  {journal} {Phys. Rev. A}\ }\textbf {\bibinfo {volume} {99}},\ \bibinfo
  {pages} {040502(R)} (\bibinfo {year} {2019})}\BibitemShut {NoStop}%
\bibitem [{\citenamefont {Pickering}\ \emph {et~al.}(2011)\citenamefont
  {Pickering}, \citenamefont {Blackwell-Whitehead}, \citenamefont {Thorne},
  \citenamefont {Ruffoni},\ and\ \citenamefont {Holmes}}]{Pickering11}%
  \BibitemOpen
  \bibfield  {author} {\bibinfo {author} {\bibfnamefont {J.}~\bibnamefont
  {Pickering}}, \bibinfo {author} {\bibfnamefont {R.}~\bibnamefont
  {Blackwell-Whitehead}}, \bibinfo {author} {\bibfnamefont {A.}~\bibnamefont
  {Thorne}}, \bibinfo {author} {\bibfnamefont {M.}~\bibnamefont {Ruffoni}},\
  and\ \bibinfo {author} {\bibfnamefont {C.}~\bibnamefont {Holmes}},\
  }\bibfield  {title} {\bibinfo {title} {Laboratory measurements of oscillator
  strengths and their astrophysical applications},\ }\href
  {https://doi.org/10.1139/p11-044} {\bibfield  {journal} {\bibinfo  {journal}
  {Canadian Journal of Physics}\ }\textbf {\bibinfo {volume} {89}},\ \bibinfo
  {pages} {387} (\bibinfo {year} {2011})}\BibitemShut {NoStop}%
\bibitem [{\citenamefont {Kandula}\ \emph {et~al.}(2010)\citenamefont
  {Kandula}, \citenamefont {Gohle}, \citenamefont {Pinkert}, \citenamefont
  {Ubachs},\ and\ \citenamefont {Eikema}}]{PhysRevLett.105.063001}%
  \BibitemOpen
  \bibfield  {author} {\bibinfo {author} {\bibfnamefont {D.~Z.}\ \bibnamefont
  {Kandula}}, \bibinfo {author} {\bibfnamefont {C.}~\bibnamefont {Gohle}},
  \bibinfo {author} {\bibfnamefont {T.~J.}\ \bibnamefont {Pinkert}}, \bibinfo
  {author} {\bibfnamefont {W.}~\bibnamefont {Ubachs}},\ and\ \bibinfo {author}
  {\bibfnamefont {K.~S.~E.}\ \bibnamefont {Eikema}},\ }\bibfield  {title}
  {\bibinfo {title} {Extreme ultraviolet frequency comb metrology},\ }\href
  {https://doi.org/10.1103/PhysRevLett.105.063001} {\bibfield  {journal}
  {\bibinfo  {journal} {Phys. Rev. Lett.}\ }\textbf {\bibinfo {volume} {105}},\
  \bibinfo {pages} {063001} (\bibinfo {year} {2010})}\BibitemShut {NoStop}%
\bibitem [{\citenamefont {Bergeson}\ \emph {et~al.}(1998)\citenamefont
  {Bergeson}, \citenamefont {Balakrishnan}, \citenamefont {Baldwin},
  \citenamefont {Lucatorto}, \citenamefont {Marangos}, \citenamefont
  {McIlrath}, \citenamefont {O'Brian}, \citenamefont {Rolston}, \citenamefont
  {Sansonetti}, \citenamefont {Wen}, \citenamefont {Westbrook}, \citenamefont
  {Cheng},\ and\ \citenamefont {Eyler}}]{PhysRevLett.80.3475}%
  \BibitemOpen
  \bibfield  {author} {\bibinfo {author} {\bibfnamefont {S.~D.}\ \bibnamefont
  {Bergeson}}, \bibinfo {author} {\bibfnamefont {A.}~\bibnamefont
  {Balakrishnan}}, \bibinfo {author} {\bibfnamefont {K.~G.~H.}\ \bibnamefont
  {Baldwin}}, \bibinfo {author} {\bibfnamefont {T.~B.}\ \bibnamefont
  {Lucatorto}}, \bibinfo {author} {\bibfnamefont {J.~P.}\ \bibnamefont
  {Marangos}}, \bibinfo {author} {\bibfnamefont {T.~J.}\ \bibnamefont
  {McIlrath}}, \bibinfo {author} {\bibfnamefont {T.~R.}\ \bibnamefont
  {O'Brian}}, \bibinfo {author} {\bibfnamefont {S.~L.}\ \bibnamefont
  {Rolston}}, \bibinfo {author} {\bibfnamefont {C.~J.}\ \bibnamefont
  {Sansonetti}}, \bibinfo {author} {\bibfnamefont {J.}~\bibnamefont {Wen}},
  \bibinfo {author} {\bibfnamefont {N.}~\bibnamefont {Westbrook}}, \bibinfo
  {author} {\bibfnamefont {C.~H.}\ \bibnamefont {Cheng}},\ and\ \bibinfo
  {author} {\bibfnamefont {E.~E.}\ \bibnamefont {Eyler}},\ }\bibfield  {title}
  {\bibinfo {title} {Measurement of the {H}e ground state {L}amb shift via the
  two-photon $1^{1}{S}\ensuremath{-}2^{1}{S}$ transition},\ }\href
  {https://doi.org/10.1103/PhysRevLett.80.3475} {\bibfield  {journal} {\bibinfo
   {journal} {Phys. Rev. Lett.}\ }\textbf {\bibinfo {volume} {80}},\ \bibinfo
  {pages} {3475} (\bibinfo {year} {1998})}\BibitemShut {NoStop}%
\bibitem [{\citenamefont {Smiciklas}\ and\ \citenamefont
  {Shiner}(2010)}]{PhysRevLett.105.123001}%
  \BibitemOpen
  \bibfield  {author} {\bibinfo {author} {\bibfnamefont {M.}~\bibnamefont
  {Smiciklas}}\ and\ \bibinfo {author} {\bibfnamefont {D.}~\bibnamefont
  {Shiner}},\ }\bibfield  {title} {\bibinfo {title} {Determination of the fine
  structure constant using helium fine structure},\ }\href
  {https://doi.org/10.1103/PhysRevLett.105.123001} {\bibfield  {journal}
  {\bibinfo  {journal} {Phys. Rev. Lett.}\ }\textbf {\bibinfo {volume} {105}},\
  \bibinfo {pages} {123001} (\bibinfo {year} {2010})}\BibitemShut {NoStop}%
\bibitem [{\citenamefont {Kato}\ \emph {et~al.}(2018)\citenamefont {Kato},
  \citenamefont {Skinner},\ and\ \citenamefont
  {Hessels}}]{PhysRevLett.121.143002}%
  \BibitemOpen
  \bibfield  {author} {\bibinfo {author} {\bibfnamefont {K.}~\bibnamefont
  {Kato}}, \bibinfo {author} {\bibfnamefont {T.~D.~G.}\ \bibnamefont
  {Skinner}},\ and\ \bibinfo {author} {\bibfnamefont {E.~A.}\ \bibnamefont
  {Hessels}},\ }\bibfield  {title} {\bibinfo {title} {Ultrahigh-precision
  measurement of the $n=2$ triplet $p$ fine structure of atomic helium using
  frequency-offset separated oscillatory fields},\ }\href
  {https://doi.org/10.1103/PhysRevLett.121.143002} {\bibfield  {journal}
  {\bibinfo  {journal} {Phys. Rev. Lett.}\ }\textbf {\bibinfo {volume} {121}},\
  \bibinfo {pages} {143002} (\bibinfo {year} {2018})}\BibitemShut {NoStop}%
\bibitem [{\citenamefont {Rengelink}\ \emph {et~al.}(2018)\citenamefont
  {Rengelink}, \citenamefont {van~der Werf}, \citenamefont {Notermans},
  \citenamefont {Jannin}, \citenamefont {Eikema}, \citenamefont {Hoogerland},\
  and\ \citenamefont {Vassen}}]{Rengelink2018}%
  \BibitemOpen
  \bibfield  {author} {\bibinfo {author} {\bibfnamefont {R.~J.}\ \bibnamefont
  {Rengelink}}, \bibinfo {author} {\bibfnamefont {Y.}~\bibnamefont {van~der
  Werf}}, \bibinfo {author} {\bibfnamefont {R.~P. M. J.~W.}\ \bibnamefont
  {Notermans}}, \bibinfo {author} {\bibfnamefont {R.}~\bibnamefont {Jannin}},
  \bibinfo {author} {\bibfnamefont {K.~S.~E.}\ \bibnamefont {Eikema}}, \bibinfo
  {author} {\bibfnamefont {M.~D.}\ \bibnamefont {Hoogerland}},\ and\ \bibinfo
  {author} {\bibfnamefont {W.}~\bibnamefont {Vassen}},\ }\bibfield  {title}
  {\bibinfo {title} {Precision spectroscopy of helium in a magic wavelength
  optical dipole trap},\ }\href {https://doi.org/10.1038/s41567-018-0242-5}
  {\bibfield  {journal} {\bibinfo  {journal} {Nature Physics}\ }\textbf
  {\bibinfo {volume} {14}},\ \bibinfo {pages} {1132} (\bibinfo {year}
  {2018})}\BibitemShut {NoStop}%
\bibitem [{\citenamefont {Cancio~Pastor}\ \emph {et~al.}(2012)\citenamefont
  {Cancio~Pastor}, \citenamefont {Consolino}, \citenamefont {Giusfredi},
  \citenamefont {De~Natale}, \citenamefont {Inguscio}, \citenamefont
  {Yerokhin},\ and\ \citenamefont {Pachucki}}]{PhysRevLett.108.143001}%
  \BibitemOpen
  \bibfield  {author} {\bibinfo {author} {\bibfnamefont {P.}~\bibnamefont
  {Cancio~Pastor}}, \bibinfo {author} {\bibfnamefont {L.}~\bibnamefont
  {Consolino}}, \bibinfo {author} {\bibfnamefont {G.}~\bibnamefont
  {Giusfredi}}, \bibinfo {author} {\bibfnamefont {P.}~\bibnamefont
  {De~Natale}}, \bibinfo {author} {\bibfnamefont {M.}~\bibnamefont {Inguscio}},
  \bibinfo {author} {\bibfnamefont {V.~A.}\ \bibnamefont {Yerokhin}},\ and\
  \bibinfo {author} {\bibfnamefont {K.}~\bibnamefont {Pachucki}},\ }\bibfield
  {title} {\bibinfo {title} {Frequency metrology of helium around 1083 nm and
  determination of the nuclear charge radius},\ }\href
  {https://doi.org/10.1103/PhysRevLett.108.143001} {\bibfield  {journal}
  {\bibinfo  {journal} {Phys. Rev. Lett.}\ }\textbf {\bibinfo {volume} {108}},\
  \bibinfo {pages} {143001} (\bibinfo {year} {2012})}\BibitemShut {NoStop}%
\bibitem [{\citenamefont {Drake}\ and\ \citenamefont
  {Morton}(2007)}]{Drake_2007}%
  \BibitemOpen
  \bibfield  {author} {\bibinfo {author} {\bibfnamefont {G.~W.~F.}\
  \bibnamefont {Drake}}\ and\ \bibinfo {author} {\bibfnamefont {D.~C.}\
  \bibnamefont {Morton}},\ }\bibfield  {title} {\bibinfo {title} {A multiplet
  table for neutral helium (\(^4\){H}e {I}) with transition rates},\ }\href
  {https://doi.org/10.1086/512239} {\bibfield  {journal} {\bibinfo  {journal}
  {The Astrophysical Journal Supplement Series}\ }\textbf {\bibinfo {volume}
  {170}},\ \bibinfo {pages} {251} (\bibinfo {year} {2007})}\BibitemShut
  {NoStop}%
\bibitem [{\citenamefont {Notermans}\ and\ \citenamefont
  {Vassen}(2014)}]{PhysRevLett.112.253002}%
  \BibitemOpen
  \bibfield  {author} {\bibinfo {author} {\bibfnamefont {R.~P. M. J.~W.}\
  \bibnamefont {Notermans}}\ and\ \bibinfo {author} {\bibfnamefont
  {W.}~\bibnamefont {Vassen}},\ }\bibfield  {title} {\bibinfo {title}
  {High-precision spectroscopy of the forbidden $2\text{
  }^{3}{S}_{1}\ensuremath{\rightarrow}2\text{ }^{1}{P}_{1}$ transition in
  quantum degenerate metastable helium},\ }\href
  {https://doi.org/10.1103/PhysRevLett.112.253002} {\bibfield  {journal}
  {\bibinfo  {journal} {Phys. Rev. Lett.}\ }\textbf {\bibinfo {volume} {112}},\
  \bibinfo {pages} {253002} (\bibinfo {year} {2014})}\BibitemShut {NoStop}%
\bibitem [{\citenamefont {Baklanov}\ and\ \citenamefont
  {Denisov}(1997)}]{ISI:000071951300016}%
  \BibitemOpen
  \bibfield  {author} {\bibinfo {author} {\bibfnamefont {E.~V.}\ \bibnamefont
  {Baklanov}}\ and\ \bibinfo {author} {\bibfnamefont {A.~V.}\ \bibnamefont
  {Denisov}},\ }\bibfield  {title} {\bibinfo {title} {Probability of the
  \(2^{1\!}{S}_0\, \)\textemdash\(\, 2^{3\!}{S}_1\) forbidden transition in the
  helium atom},\ }\href {https://doi.org/10.1070/qe1997v027n05abeh000963}
  {\bibfield  {journal} {\bibinfo  {journal} {Quantum Electronics}\ }\textbf
  {\bibinfo {volume} {27}},\ \bibinfo {pages} {463} (\bibinfo {year}
  {1997})}\BibitemShut {NoStop}%
\bibitem [{\citenamefont {Lin}\ \emph {et~al.}(1977)\citenamefont {Lin},
  \citenamefont {Johnson},\ and\ \citenamefont {Dalgarno}}]{PhysRevA.15.154}%
  \BibitemOpen
  \bibfield  {author} {\bibinfo {author} {\bibfnamefont {C.~D.}\ \bibnamefont
  {Lin}}, \bibinfo {author} {\bibfnamefont {W.~R.}\ \bibnamefont {Johnson}},\
  and\ \bibinfo {author} {\bibfnamefont {A.}~\bibnamefont {Dalgarno}},\
  }\bibfield  {title} {\bibinfo {title} {Radiative decays of the $n=2$ states
  of he-like ions},\ }\href {https://doi.org/10.1103/PhysRevA.15.154}
  {\bibfield  {journal} {\bibinfo  {journal} {Phys. Rev. A}\ }\textbf {\bibinfo
  {volume} {15}},\ \bibinfo {pages} {154} (\bibinfo {year} {1977})}\BibitemShut
  {NoStop}%
\bibitem [{\citenamefont {van Rooij}\ \emph {et~al.}(2011)\citenamefont {van
  Rooij}, \citenamefont {Borbely}, \citenamefont {Simonet}, \citenamefont
  {Hoogerland}, \citenamefont {Eikema}, \citenamefont {Rozendaal},\ and\
  \citenamefont {Vassen}}]{vanRooij196}%
  \BibitemOpen
  \bibfield  {author} {\bibinfo {author} {\bibfnamefont {R.}~\bibnamefont {van
  Rooij}}, \bibinfo {author} {\bibfnamefont {J.~S.}\ \bibnamefont {Borbely}},
  \bibinfo {author} {\bibfnamefont {J.}~\bibnamefont {Simonet}}, \bibinfo
  {author} {\bibfnamefont {M.~D.}\ \bibnamefont {Hoogerland}}, \bibinfo
  {author} {\bibfnamefont {K.~S.~E.}\ \bibnamefont {Eikema}}, \bibinfo {author}
  {\bibfnamefont {R.~A.}\ \bibnamefont {Rozendaal}},\ and\ \bibinfo {author}
  {\bibfnamefont {W.}~\bibnamefont {Vassen}},\ }\bibfield  {title} {\bibinfo
  {title} {Frequency metrology in quantum degenerate helium: Direct measurement
  of the $2^{3}{S}_{1} \rightarrow 2^{1}{S}_{0}$ transition},\ }\href
  {https://doi.org/10.1126/science.1205163} {\bibfield  {journal} {\bibinfo
  {journal} {Science}\ }\textbf {\bibinfo {volume} {333}},\ \bibinfo {pages}
  {196} (\bibinfo {year} {2011})}\BibitemShut {NoStop}%
\bibitem [{\citenamefont {Drake}(2006)}]{Drake2006}%
  \BibitemOpen
  \bibfield  {author} {\bibinfo {author} {\bibfnamefont {G.~W.~F.}\
  \bibnamefont {Drake}},\ }\bibinfo {title} {Springer handbook of atomic,
  molecular, and optical physics}\ (\bibinfo {year} {2006})\ Chap.\ \bibinfo
  {chapter} {High Precision Calculations for Helium}, pp.\ \bibinfo {pages}
  {199--219}\BibitemShut {NoStop}%
\bibitem [{\citenamefont {Derevianko}\ \emph {et~al.}(1998)\citenamefont
  {Derevianko}, \citenamefont {Savukov}, \citenamefont {Johnson},\ and\
  \citenamefont {Plante}}]{PhysRevA.58.4453}%
  \BibitemOpen
  \bibfield  {author} {\bibinfo {author} {\bibfnamefont {A.}~\bibnamefont
  {Derevianko}}, \bibinfo {author} {\bibfnamefont {I.~M.}\ \bibnamefont
  {Savukov}}, \bibinfo {author} {\bibfnamefont {W.~R.}\ \bibnamefont
  {Johnson}},\ and\ \bibinfo {author} {\bibfnamefont {D.~R.}\ \bibnamefont
  {Plante}},\ }\bibfield  {title} {\bibinfo {title} {Negative-energy
  contributions to transition amplitudes in heliumlike ions},\ }\href
  {https://doi.org/10.1103/PhysRevA.58.4453} {\bibfield  {journal} {\bibinfo
  {journal} {Phys. Rev. A}\ }\textbf {\bibinfo {volume} {58}},\ \bibinfo
  {pages} {4453} (\bibinfo {year} {1998})}\BibitemShut {NoStop}%
\bibitem [{\citenamefont {\L{}ach}\ and\ \citenamefont
  {Pachucki}(2001)}]{PhysRevA.64.042510}%
  \BibitemOpen
  \bibfield  {author} {\bibinfo {author} {\bibfnamefont {G.}~\bibnamefont
  {\L{}ach}}\ and\ \bibinfo {author} {\bibfnamefont {K.}~\bibnamefont
  {Pachucki}},\ }\bibfield  {title} {\bibinfo {title} {Forbidden transitions in
  the helium atom},\ }\href {https://doi.org/10.1103/PhysRevA.64.042510}
  {\bibfield  {journal} {\bibinfo  {journal} {Phys. Rev. A}\ }\textbf {\bibinfo
  {volume} {64}},\ \bibinfo {pages} {042510} (\bibinfo {year}
  {2001})}\BibitemShut {NoStop}%
\bibitem [{\citenamefont {Bi\'emont}\ and\ \citenamefont
  {Quinet}(1998)}]{PhysRevLett.81.3345}%
  \BibitemOpen
  \bibfield  {author} {\bibinfo {author} {\bibfnamefont {E.}~\bibnamefont
  {Bi\'emont}}\ and\ \bibinfo {author} {\bibfnamefont {P.}~\bibnamefont
  {Quinet}},\ }\bibfield  {title} {\bibinfo {title} {Theoretical study of the
  $4{\mathit{f}}^{14}6\mathit{s}$
  ${}^{2}{{\mathit{S}}}_{1/2}--4{\mathit{f}}^{13}6{\mathit{s}}^{2}$
  ${}^{2}{{\mathit{F}}}_{7/2}^{0}$ $\mathit{E}3$ transition in {Y}b {I}{I}},\
  }\href {https://doi.org/10.1103/PhysRevLett.81.3345} {\bibfield  {journal}
  {\bibinfo  {journal} {Phys. Rev. Lett.}\ }\textbf {\bibinfo {volume} {81}},\
  \bibinfo {pages} {3345} (\bibinfo {year} {1998})}\BibitemShut {NoStop}%
\bibitem [{\citenamefont {Roberts}\ \emph {et~al.}(1997)\citenamefont
  {Roberts}, \citenamefont {Taylor}, \citenamefont {Barwood}, \citenamefont
  {Gill}, \citenamefont {Klein},\ and\ \citenamefont
  {Rowley}}]{PhysRevLett.78.1876}%
  \BibitemOpen
  \bibfield  {author} {\bibinfo {author} {\bibfnamefont {M.}~\bibnamefont
  {Roberts}}, \bibinfo {author} {\bibfnamefont {P.}~\bibnamefont {Taylor}},
  \bibinfo {author} {\bibfnamefont {G.~P.}\ \bibnamefont {Barwood}}, \bibinfo
  {author} {\bibfnamefont {P.}~\bibnamefont {Gill}}, \bibinfo {author}
  {\bibfnamefont {H.~A.}\ \bibnamefont {Klein}},\ and\ \bibinfo {author}
  {\bibfnamefont {W.~R.~C.}\ \bibnamefont {Rowley}},\ }\bibfield  {title}
  {\bibinfo {title} {Observation of an electric octupole transition in a single
  ion},\ }\href {https://doi.org/10.1103/PhysRevLett.78.1876} {\bibfield
  {journal} {\bibinfo  {journal} {Phys. Rev. Lett.}\ }\textbf {\bibinfo
  {volume} {78}},\ \bibinfo {pages} {1876} (\bibinfo {year}
  {1997})}\BibitemShut {NoStop}%
\bibitem [{\citenamefont {Wan}\ \emph {et~al.}(2014)\citenamefont {Wan},
  \citenamefont {Gebert}, \citenamefont {W\"{u}bbena}, \citenamefont
  {Scharnhorst}, \citenamefont {Amairi}, \citenamefont {Leroux}, \citenamefont
  {Hemmerling}, \citenamefont {L\"{o}rch}, \citenamefont {Hammerer},\ and\
  \citenamefont {Schmidt}}]{Wan2014}%
  \BibitemOpen
  \bibfield  {author} {\bibinfo {author} {\bibfnamefont {Y.}~\bibnamefont
  {Wan}}, \bibinfo {author} {\bibfnamefont {F.}~\bibnamefont {Gebert}},
  \bibinfo {author} {\bibfnamefont {J.~B.}\ \bibnamefont {W\"{u}bbena}},
  \bibinfo {author} {\bibfnamefont {N.}~\bibnamefont {Scharnhorst}}, \bibinfo
  {author} {\bibfnamefont {S.}~\bibnamefont {Amairi}}, \bibinfo {author}
  {\bibfnamefont {I.~D.}\ \bibnamefont {Leroux}}, \bibinfo {author}
  {\bibfnamefont {B.}~\bibnamefont {Hemmerling}}, \bibinfo {author}
  {\bibfnamefont {N.}~\bibnamefont {L\"{o}rch}}, \bibinfo {author}
  {\bibfnamefont {K.}~\bibnamefont {Hammerer}},\ and\ \bibinfo {author}
  {\bibfnamefont {P.~O.}\ \bibnamefont {Schmidt}},\ }\bibfield  {title}
  {\bibinfo {title} {Precision spectroscopy by photon-recoil signal
  amplification},\ }\href {https://doi.org/10.1038/ncomms4096} {\bibfield
  {journal} {\bibinfo  {journal} {Nature Communications}\ }\textbf {\bibinfo
  {volume} {5}} (\bibinfo {year} {2014})}\BibitemShut {NoStop}%
\bibitem [{\citenamefont {Gebert}\ \emph {et~al.}(2015)\citenamefont {Gebert},
  \citenamefont {Wan}, \citenamefont {Wolf}, \citenamefont {Angstmann},
  \citenamefont {Berengut},\ and\ \citenamefont
  {Schmidt}}]{PhysRevLett.115.053003}%
  \BibitemOpen
  \bibfield  {author} {\bibinfo {author} {\bibfnamefont {F.}~\bibnamefont
  {Gebert}}, \bibinfo {author} {\bibfnamefont {Y.}~\bibnamefont {Wan}},
  \bibinfo {author} {\bibfnamefont {F.}~\bibnamefont {Wolf}}, \bibinfo {author}
  {\bibfnamefont {C.~N.}\ \bibnamefont {Angstmann}}, \bibinfo {author}
  {\bibfnamefont {J.~C.}\ \bibnamefont {Berengut}},\ and\ \bibinfo {author}
  {\bibfnamefont {P.~O.}\ \bibnamefont {Schmidt}},\ }\bibfield  {title}
  {\bibinfo {title} {Precision isotope shift measurements in calcium ions using
  quantum logic detection schemes},\ }\href
  {https://doi.org/10.1103/PhysRevLett.115.053003} {\bibfield  {journal}
  {\bibinfo  {journal} {Phys. Rev. Lett.}\ }\textbf {\bibinfo {volume} {115}},\
  \bibinfo {pages} {053003} (\bibinfo {year} {2015})}\BibitemShut {NoStop}%
\bibitem [{\citenamefont {Guggemos}\ \emph {et~al.}(2019)\citenamefont
  {Guggemos}, \citenamefont {Guevara-Bertsch}, \citenamefont {Heinrich},
  \citenamefont {Herrera-Sancho}, \citenamefont {Colombe}, \citenamefont
  {Blatt},\ and\ \citenamefont {Roos}}]{Guggemos_2019}%
  \BibitemOpen
  \bibfield  {author} {\bibinfo {author} {\bibfnamefont {M.}~\bibnamefont
  {Guggemos}}, \bibinfo {author} {\bibfnamefont {M.}~\bibnamefont
  {Guevara-Bertsch}}, \bibinfo {author} {\bibfnamefont {D.}~\bibnamefont
  {Heinrich}}, \bibinfo {author} {\bibfnamefont {O.~A.}\ \bibnamefont
  {Herrera-Sancho}}, \bibinfo {author} {\bibfnamefont {Y.}~\bibnamefont
  {Colombe}}, \bibinfo {author} {\bibfnamefont {R.}~\bibnamefont {Blatt}},\
  and\ \bibinfo {author} {\bibfnamefont {C.~F.}\ \bibnamefont {Roos}},\
  }\bibfield  {title} {\bibinfo {title} {Frequency measurement of the
  ${}^{1}{{\rm{S}}}_{0},f=5/2\,\leftrightarrow {}^{3}{{\rm{P}}}_{1},f=7/2$
  transition of $^{27}${A}l$^{+}$ via quantum logic spectroscopy with
  $^{40}${C}a$^{+}$},\ }\href {https://doi.org/10.1088/1367-2630/ab447a}
  {\bibfield  {journal} {\bibinfo  {journal} {New Journal of Physics}\ }\textbf
  {\bibinfo {volume} {21}},\ \bibinfo {pages} {103003} (\bibinfo {year}
  {2019})}\BibitemShut {NoStop}%
\bibitem [{det()}]{detection_note}%
  \BibitemOpen
  \href@noop {} {}\bibinfo {note} {This detection method only works for atoms
  with a high internal energy, such as He\(^*\) atoms or other excited noble
  gases; for other atomic species lacking this property one would have to use a
  different, though comparable, detection method such as absorption imaging or
  continuous light sheet detection.}\BibitemShut {Stop}%
\bibitem [{\citenamefont {Hodgman}\ \emph
  {et~al.}(2009{\natexlab{a}})\citenamefont {Hodgman}, \citenamefont {Dall},
  \citenamefont {Byron}, \citenamefont {Baldwin}, \citenamefont {Buckman},\
  and\ \citenamefont {Truscott}}]{PhysRevLett.103.053002}%
  \BibitemOpen
  \bibfield  {author} {\bibinfo {author} {\bibfnamefont {S.~S.}\ \bibnamefont
  {Hodgman}}, \bibinfo {author} {\bibfnamefont {R.~G.}\ \bibnamefont {Dall}},
  \bibinfo {author} {\bibfnamefont {L.~J.}\ \bibnamefont {Byron}}, \bibinfo
  {author} {\bibfnamefont {K.~G.~H.}\ \bibnamefont {Baldwin}}, \bibinfo
  {author} {\bibfnamefont {S.~J.}\ \bibnamefont {Buckman}},\ and\ \bibinfo
  {author} {\bibfnamefont {A.~G.}\ \bibnamefont {Truscott}},\ }\bibfield
  {title} {\bibinfo {title} {Metastable helium: A new determination of the
  longest atomic excited-state lifetime},\ }\href
  {https://doi.org/10.1103/PhysRevLett.103.053002} {\bibfield  {journal}
  {\bibinfo  {journal} {Phys. Rev. Lett.}\ }\textbf {\bibinfo {volume} {103}},\
  \bibinfo {pages} {053002} (\bibinfo {year} {2009}{\natexlab{a}})}\BibitemShut
  {NoStop}%
\bibitem [{\citenamefont {Dall}\ and\ \citenamefont
  {Truscott}(2007)}]{Dall2007}%
  \BibitemOpen
  \bibfield  {author} {\bibinfo {author} {\bibfnamefont {R.}~\bibnamefont
  {Dall}}\ and\ \bibinfo {author} {\bibfnamefont {A.}~\bibnamefont
  {Truscott}},\ }\bibfield  {title} {\bibinfo {title}
  {Bose{\textendash}{E}instein condensation of metastable helium in a bi-planar
  quadrupole {I}offe configuration trap},\ }\href
  {https://doi.org/10.1016/j.optcom.2006.09.031} {\bibfield  {journal}
  {\bibinfo  {journal} {Optics Communications}\ }\textbf {\bibinfo {volume}
  {270}},\ \bibinfo {pages} {255} (\bibinfo {year} {2007})}\BibitemShut
  {NoStop}%
\bibitem [{\citenamefont {Manning}\ \emph {et~al.}(2010)\citenamefont
  {Manning}, \citenamefont {Hodgman}, \citenamefont {Dall}, \citenamefont
  {Johnsson},\ and\ \citenamefont {Truscott}}]{Manning:10}%
  \BibitemOpen
  \bibfield  {author} {\bibinfo {author} {\bibfnamefont {A.~G.}\ \bibnamefont
  {Manning}}, \bibinfo {author} {\bibfnamefont {S.~S.}\ \bibnamefont
  {Hodgman}}, \bibinfo {author} {\bibfnamefont {R.~G.}\ \bibnamefont {Dall}},
  \bibinfo {author} {\bibfnamefont {M.~T.}\ \bibnamefont {Johnsson}},\ and\
  \bibinfo {author} {\bibfnamefont {A.~G.}\ \bibnamefont {Truscott}},\
  }\bibfield  {title} {\bibinfo {title} {The {H}anbury {B}rown-{T}wiss effect
  in a pulsed atom laser},\ }\href {https://doi.org/10.1364/OE.18.018712}
  {\bibfield  {journal} {\bibinfo  {journal} {Opt. Express}\ }\textbf {\bibinfo
  {volume} {18}},\ \bibinfo {pages} {18712} (\bibinfo {year}
  {2010})}\BibitemShut {NoStop}%
\bibitem [{SOM()}]{SOMs}%
  \BibitemOpen
  \href {https://link.aps.org/doi/10.1103/PhysRevLett.120.103201} {}\bibinfo
  {note} {See Supplementary Material for further details, which includes
  Refs.~\cite{pmid9900545,wstechnical,Kondratjev2010,Julienne:89,PhysRevLett.81.3807,PhysRevLett.81.3807,PhysRevLett.96.023203,PhysRevLett.93.090409,PhysRevA.48.546,PhysRevA.64.042710,TALU200183,PhysRevA.80.062718,PhysRevA.82.042502,1964JChEd..41R.590M,PhysRevA.88.052515,astapenko2013polarization,LeKien2013,PhysRevA.73.043416,CohenTannoudji1996,PhysRevA.73.043416,Fendel:07,Wu:13,NISTASD,Huber1986,doi:10.1119/1.3417868,Krainov2019,metcalf1999laser}.}\BibitemShut
  {Stop}%
\bibitem [{\citenamefont {Dall}\ \emph {et~al.}(2008)\citenamefont {Dall},
  \citenamefont {Baldwin}, \citenamefont {Byron},\ and\ \citenamefont
  {Truscott}}]{PhysRevLett.100.023001}%
  \BibitemOpen
  \bibfield  {author} {\bibinfo {author} {\bibfnamefont {R.~G.}\ \bibnamefont
  {Dall}}, \bibinfo {author} {\bibfnamefont {K.~G.~H.}\ \bibnamefont
  {Baldwin}}, \bibinfo {author} {\bibfnamefont {L.~J.}\ \bibnamefont {Byron}},\
  and\ \bibinfo {author} {\bibfnamefont {A.~G.}\ \bibnamefont {Truscott}},\
  }\bibfield  {title} {\bibinfo {title} {Experimental determination of the
  helium $2\text{ }^{3}{P}_{1}--1\text{ }^{1}{S}_{0}$ transition rate},\ }\href
  {https://doi.org/10.1103/PhysRevLett.100.023001} {\bibfield  {journal}
  {\bibinfo  {journal} {Phys. Rev. Lett.}\ }\textbf {\bibinfo {volume} {100}},\
  \bibinfo {pages} {023001} (\bibinfo {year} {2008})}\BibitemShut {NoStop}%
\bibitem [{\citenamefont {Hodgman}\ \emph
  {et~al.}(2009{\natexlab{b}})\citenamefont {Hodgman}, \citenamefont {Dall},
  \citenamefont {Baldwin},\ and\ \citenamefont {Truscott}}]{Hodgman2009a}%
  \BibitemOpen
  \bibfield  {author} {\bibinfo {author} {\bibfnamefont {S.~S.}\ \bibnamefont
  {Hodgman}}, \bibinfo {author} {\bibfnamefont {R.~G.}\ \bibnamefont {Dall}},
  \bibinfo {author} {\bibfnamefont {K.~G.~H.}\ \bibnamefont {Baldwin}},\ and\
  \bibinfo {author} {\bibfnamefont {A.~G.}\ \bibnamefont {Truscott}},\
  }\bibfield  {title} {\bibinfo {title} {Complete ground-state transition rates
  for the helium $2\text{ }^{3}{P}$ manifold},\ }\href
  {https://doi.org/10.1103/PhysRevA.80.044501} {\bibfield  {journal} {\bibinfo
  {journal} {Phys. Rev. A}\ }\textbf {\bibinfo {volume} {80}},\ \bibinfo
  {pages} {044501} (\bibinfo {year} {2009}{\natexlab{b}})}\BibitemShut
  {NoStop}%
\bibitem [{sta()}]{standard_error_note}%
  \BibitemOpen
  \href@noop {} {}\bibinfo {note} {Vertical bars indicate standard error in
  signal of the respective sample, and horizontal bars are the quadrature sum
  of the standard error in frequency and the standard error of wavemeter
  calibrations}\BibitemShut {NoStop}%
\bibitem [{\citenamefont {Henson}\ \emph {et~al.}(2018)\citenamefont {Henson},
  \citenamefont {Yue}, \citenamefont {Hodgman}, \citenamefont {Shin},
  \citenamefont {Smirnov}, \citenamefont {Ostrovskaya}, \citenamefont {Guan},\
  and\ \citenamefont {Truscott}}]{Henson2018}%
  \BibitemOpen
  \bibfield  {author} {\bibinfo {author} {\bibfnamefont {B.~M.}\ \bibnamefont
  {Henson}}, \bibinfo {author} {\bibfnamefont {X.}~\bibnamefont {Yue}},
  \bibinfo {author} {\bibfnamefont {S.~S.}\ \bibnamefont {Hodgman}}, \bibinfo
  {author} {\bibfnamefont {D.~K.}\ \bibnamefont {Shin}}, \bibinfo {author}
  {\bibfnamefont {L.~A.}\ \bibnamefont {Smirnov}}, \bibinfo {author}
  {\bibfnamefont {E.~A.}\ \bibnamefont {Ostrovskaya}}, \bibinfo {author}
  {\bibfnamefont {X.~W.}\ \bibnamefont {Guan}},\ and\ \bibinfo {author}
  {\bibfnamefont {A.~G.}\ \bibnamefont {Truscott}},\ }\bibfield  {title}
  {\bibinfo {title} {Bogoliubov-cherenkov radiation in an atom laser},\ }\href
  {https://doi.org/10.1103/PhysRevA.97.063601} {\bibfield  {journal} {\bibinfo
  {journal} {Phys. Rev. A}\ }\textbf {\bibinfo {volume} {97}},\ \bibinfo
  {pages} {063601} (\bibinfo {year} {2018})}\BibitemShut {NoStop}%
\bibitem [{\citenamefont {Yavin}\ \emph {et~al.}(2002)\citenamefont {Yavin},
  \citenamefont {Weel}, \citenamefont {Andreyuk},\ and\ \citenamefont
  {Kumarakrishnan}}]{Yavin2002}%
  \BibitemOpen
  \bibfield  {author} {\bibinfo {author} {\bibfnamefont {I.}~\bibnamefont
  {Yavin}}, \bibinfo {author} {\bibfnamefont {M.}~\bibnamefont {Weel}},
  \bibinfo {author} {\bibfnamefont {A.}~\bibnamefont {Andreyuk}},\ and\
  \bibinfo {author} {\bibfnamefont {A.}~\bibnamefont {Kumarakrishnan}},\
  }\bibfield  {title} {\bibinfo {title} {A calculation of the time-of-flight
  distribution of trapped atoms},\ }\href {https://doi.org/10.1119/1.1424266}
  {\bibfield  {journal} {\bibinfo  {journal} {American Journal of Physics}\
  }\textbf {\bibinfo {volume} {70}},\ \bibinfo {pages} {149} (\bibinfo {year}
  {2002})}\BibitemShut {NoStop}%
\bibitem [{\citenamefont {Zheng}\ \emph {et~al.}(2017)\citenamefont {Zheng},
  \citenamefont {Sun}, \citenamefont {Chen}, \citenamefont {Jiang},
  \citenamefont {Pachucki},\ and\ \citenamefont {Hu}}]{PhysRevLett.119.263002}%
  \BibitemOpen
  \bibfield  {author} {\bibinfo {author} {\bibfnamefont {X.}~\bibnamefont
  {Zheng}}, \bibinfo {author} {\bibfnamefont {Y.~R.}\ \bibnamefont {Sun}},
  \bibinfo {author} {\bibfnamefont {J.-J.}\ \bibnamefont {Chen}}, \bibinfo
  {author} {\bibfnamefont {W.}~\bibnamefont {Jiang}}, \bibinfo {author}
  {\bibfnamefont {K.}~\bibnamefont {Pachucki}},\ and\ \bibinfo {author}
  {\bibfnamefont {S.-M.}\ \bibnamefont {Hu}},\ }\bibfield  {title} {\bibinfo
  {title} {Measurement of the frequency of the $2\text{
  }^{3}{S}\ensuremath{-}2\text{ }^{3}{P}$ transition of $^{4}\mathrm{He}$},\
  }\href {https://doi.org/10.1103/PhysRevLett.119.263002} {\bibfield  {journal}
  {\bibinfo  {journal} {Phys. Rev. Lett.}\ }\textbf {\bibinfo {volume} {119}},\
  \bibinfo {pages} {263002} (\bibinfo {year} {2017})}\BibitemShut {NoStop}%
\bibitem [{\citenamefont {Vassen}\ \emph {et~al.}(2012)\citenamefont {Vassen},
  \citenamefont {Cohen-Tannoudji}, \citenamefont {Leduc}, \citenamefont
  {Boiron}, \citenamefont {Westbrook}, \citenamefont {Truscott}, \citenamefont
  {Baldwin}, \citenamefont {Birkl}, \citenamefont {Cancio},\ and\ \citenamefont
  {Trippenbach}}]{RevModPhys.84.175}%
  \BibitemOpen
  \bibfield  {author} {\bibinfo {author} {\bibfnamefont {W.}~\bibnamefont
  {Vassen}}, \bibinfo {author} {\bibfnamefont {C.}~\bibnamefont
  {Cohen-Tannoudji}}, \bibinfo {author} {\bibfnamefont {M.}~\bibnamefont
  {Leduc}}, \bibinfo {author} {\bibfnamefont {D.}~\bibnamefont {Boiron}},
  \bibinfo {author} {\bibfnamefont {C.~I.}\ \bibnamefont {Westbrook}}, \bibinfo
  {author} {\bibfnamefont {A.}~\bibnamefont {Truscott}}, \bibinfo {author}
  {\bibfnamefont {K.}~\bibnamefont {Baldwin}}, \bibinfo {author} {\bibfnamefont
  {G.}~\bibnamefont {Birkl}}, \bibinfo {author} {\bibfnamefont
  {P.}~\bibnamefont {Cancio}},\ and\ \bibinfo {author} {\bibfnamefont
  {M.}~\bibnamefont {Trippenbach}},\ }\bibfield  {title} {\bibinfo {title}
  {Cold and trapped metastable noble gases},\ }\href
  {https://doi.org/10.1103/RevModPhys.84.175} {\bibfield  {journal} {\bibinfo
  {journal} {Rev. Mod. Phys.}\ }\textbf {\bibinfo {volume} {84}},\ \bibinfo
  {pages} {175} (\bibinfo {year} {2012})}\BibitemShut {NoStop}%
\bibitem [{\citenamefont {Tanner}\ and\ \citenamefont
  {Wieman}(1988)}]{pmid9900545}%
  \BibitemOpen
  \bibfield  {author} {\bibinfo {author} {\bibfnamefont {C.~E.}\ \bibnamefont
  {Tanner}}\ and\ \bibinfo {author} {\bibfnamefont {C.}~\bibnamefont
  {Wieman}},\ }\bibfield  {title} {\bibinfo {title} {Precision measurement of
  the hyperfine structure of the $^{133}\mathrm{Cs}$ 6${P}_{3/2}$ state},\
  }\href {https://doi.org/10.1103/PhysRevA.38.1616} {\bibfield  {journal}
  {\bibinfo  {journal} {Phys. Rev. A}\ }\textbf {\bibinfo {volume} {38}},\
  \bibinfo {pages} {1616} (\bibinfo {year} {1988})}\BibitemShut {NoStop}%
\bibitem [{wst(2019)}]{wstechnical}%
  \BibitemOpen
  \href
  {https://www.highfinesse.com/en/wavelengthmeter/wavelengthmeter-further-information/technical-information-wavelengthmeter-ws8-2.pdf}
  {\emph {\bibinfo {title} {Technical Information wavelengthmeter WS8-2}}},\
  \bibinfo {organization} {HighFinesse GmbH} (\bibinfo {year}
  {2019})\BibitemShut {NoStop}%
\bibitem [{\citenamefont {Kondratjev}\ \emph {et~al.}(2010)\citenamefont
  {Kondratjev}, \citenamefont {Beigman},\ and\ \citenamefont
  {Vainshtein}}]{Kondratjev2010}%
  \BibitemOpen
  \bibfield  {author} {\bibinfo {author} {\bibfnamefont {D.~A.}\ \bibnamefont
  {Kondratjev}}, \bibinfo {author} {\bibfnamefont {I.~L.}\ \bibnamefont
  {Beigman}},\ and\ \bibinfo {author} {\bibfnamefont {L.~A.}\ \bibnamefont
  {Vainshtein}},\ }\bibfield  {title} {\bibinfo {title} {Static
  polarizabilities of helium and alkali atoms, and their isoelectronic ions},\
  }\href {https://doi.org/10.1007/s10946-010-9148-0} {\bibfield  {journal}
  {\bibinfo  {journal} {Journal of Russian Laser Research}\ }\textbf {\bibinfo
  {volume} {31}},\ \bibinfo {pages} {294} (\bibinfo {year} {2010})}\BibitemShut
  {NoStop}%
\bibitem [{\citenamefont {Julienne}\ and\ \citenamefont
  {Mies}(1989)}]{Julienne:89}%
  \BibitemOpen
  \bibfield  {author} {\bibinfo {author} {\bibfnamefont {P.~S.}\ \bibnamefont
  {Julienne}}\ and\ \bibinfo {author} {\bibfnamefont {F.~H.}\ \bibnamefont
  {Mies}},\ }\bibfield  {title} {\bibinfo {title} {Collisions of ultracold
  trapped atoms},\ }\href {https://doi.org/10.1364/JOSAB.6.002257} {\bibfield
  {journal} {\bibinfo  {journal} {J. Opt. Soc. Am. B}\ }\textbf {\bibinfo
  {volume} {6}},\ \bibinfo {pages} {2257} (\bibinfo {year} {1989})}\BibitemShut
  {NoStop}%
\bibitem [{\citenamefont {Killian}\ \emph {et~al.}(1998)\citenamefont
  {Killian}, \citenamefont {Fried}, \citenamefont {Willmann}, \citenamefont
  {Landhuis}, \citenamefont {Moss}, \citenamefont {Greytak},\ and\
  \citenamefont {Kleppner}}]{PhysRevLett.81.3807}%
  \BibitemOpen
  \bibfield  {author} {\bibinfo {author} {\bibfnamefont {T.~C.}\ \bibnamefont
  {Killian}}, \bibinfo {author} {\bibfnamefont {D.~G.}\ \bibnamefont {Fried}},
  \bibinfo {author} {\bibfnamefont {L.}~\bibnamefont {Willmann}}, \bibinfo
  {author} {\bibfnamefont {D.}~\bibnamefont {Landhuis}}, \bibinfo {author}
  {\bibfnamefont {S.~C.}\ \bibnamefont {Moss}}, \bibinfo {author}
  {\bibfnamefont {T.~J.}\ \bibnamefont {Greytak}},\ and\ \bibinfo {author}
  {\bibfnamefont {D.}~\bibnamefont {Kleppner}},\ }\bibfield  {title} {\bibinfo
  {title} {Cold collision frequency shift of the $1\mathit{S}$- $2\mathit{S}$
  transition in hydrogen},\ }\href
  {https://doi.org/10.1103/PhysRevLett.81.3807} {\bibfield  {journal} {\bibinfo
   {journal} {Phys. Rev. Lett.}\ }\textbf {\bibinfo {volume} {81}},\ \bibinfo
  {pages} {3807} (\bibinfo {year} {1998})}\BibitemShut {NoStop}%
\bibitem [{\citenamefont {Moal}\ \emph {et~al.}(2006)\citenamefont {Moal},
  \citenamefont {Portier}, \citenamefont {Kim}, \citenamefont {Dugu\'e},
  \citenamefont {Rapol}, \citenamefont {Leduc},\ and\ \citenamefont
  {Cohen-Tannoudji}}]{PhysRevLett.96.023203}%
  \BibitemOpen
  \bibfield  {author} {\bibinfo {author} {\bibfnamefont {S.}~\bibnamefont
  {Moal}}, \bibinfo {author} {\bibfnamefont {M.}~\bibnamefont {Portier}},
  \bibinfo {author} {\bibfnamefont {J.}~\bibnamefont {Kim}}, \bibinfo {author}
  {\bibfnamefont {J.}~\bibnamefont {Dugu\'e}}, \bibinfo {author} {\bibfnamefont
  {U.~D.}\ \bibnamefont {Rapol}}, \bibinfo {author} {\bibfnamefont
  {M.}~\bibnamefont {Leduc}},\ and\ \bibinfo {author} {\bibfnamefont
  {C.}~\bibnamefont {Cohen-Tannoudji}},\ }\bibfield  {title} {\bibinfo {title}
  {Accurate determination of the scattering length of metastable helium atoms
  using dark resonances between atoms and exotic molecules},\ }\href
  {https://doi.org/10.1103/PhysRevLett.96.023203} {\bibfield  {journal}
  {\bibinfo  {journal} {Phys. Rev. Lett.}\ }\textbf {\bibinfo {volume} {96}},\
  \bibinfo {pages} {023203} (\bibinfo {year} {2006})}\BibitemShut {NoStop}%
\bibitem [{\citenamefont {Seidelin}\ \emph {et~al.}(2004)\citenamefont
  {Seidelin}, \citenamefont {Gomes}, \citenamefont {Hoppeler}, \citenamefont
  {Sirjean}, \citenamefont {Boiron}, \citenamefont {Aspect},\ and\
  \citenamefont {Westbrook}}]{PhysRevLett.93.090409}%
  \BibitemOpen
  \bibfield  {author} {\bibinfo {author} {\bibfnamefont {S.}~\bibnamefont
  {Seidelin}}, \bibinfo {author} {\bibfnamefont {J.~V.}\ \bibnamefont {Gomes}},
  \bibinfo {author} {\bibfnamefont {R.}~\bibnamefont {Hoppeler}}, \bibinfo
  {author} {\bibfnamefont {O.}~\bibnamefont {Sirjean}}, \bibinfo {author}
  {\bibfnamefont {D.}~\bibnamefont {Boiron}}, \bibinfo {author} {\bibfnamefont
  {A.}~\bibnamefont {Aspect}},\ and\ \bibinfo {author} {\bibfnamefont {C.~I.}\
  \bibnamefont {Westbrook}},\ }\bibfield  {title} {\bibinfo {title} {Getting
  the elastic scattering length by observing inelastic collisions in ultracold
  metastable helium atoms},\ }\href
  {https://doi.org/10.1103/PhysRevLett.93.090409} {\bibfield  {journal}
  {\bibinfo  {journal} {Phys. Rev. Lett.}\ }\textbf {\bibinfo {volume} {93}},\
  \bibinfo {pages} {090409} (\bibinfo {year} {2004})}\BibitemShut {NoStop}%
\bibitem [{\citenamefont {Gribakin}\ and\ \citenamefont
  {Flambaum}(1993)}]{PhysRevA.48.546}%
  \BibitemOpen
  \bibfield  {author} {\bibinfo {author} {\bibfnamefont {G.~F.}\ \bibnamefont
  {Gribakin}}\ and\ \bibinfo {author} {\bibfnamefont {V.~V.}\ \bibnamefont
  {Flambaum}},\ }\bibfield  {title} {\bibinfo {title} {Calculation of the
  scattering length in atomic collisions using the semiclassical
  approximation},\ }\href {https://doi.org/10.1103/PhysRevA.48.546} {\bibfield
  {journal} {\bibinfo  {journal} {Phys. Rev. A}\ }\textbf {\bibinfo {volume}
  {48}},\ \bibinfo {pages} {546} (\bibinfo {year} {1993})}\BibitemShut
  {NoStop}%
\bibitem [{\citenamefont {Leo}\ \emph {et~al.}(2001)\citenamefont {Leo},
  \citenamefont {Venturi}, \citenamefont {Whittingham},\ and\ \citenamefont
  {Babb}}]{PhysRevA.64.042710}%
  \BibitemOpen
  \bibfield  {author} {\bibinfo {author} {\bibfnamefont {P.~J.}\ \bibnamefont
  {Leo}}, \bibinfo {author} {\bibfnamefont {V.}~\bibnamefont {Venturi}},
  \bibinfo {author} {\bibfnamefont {I.~B.}\ \bibnamefont {Whittingham}},\ and\
  \bibinfo {author} {\bibfnamefont {J.~F.}\ \bibnamefont {Babb}},\ }\bibfield
  {title} {\bibinfo {title} {Ultracold collisions of metastable helium atoms},\
  }\href {https://doi.org/10.1103/PhysRevA.64.042710} {\bibfield  {journal}
  {\bibinfo  {journal} {Phys. Rev. A}\ }\textbf {\bibinfo {volume} {64}},\
  \bibinfo {pages} {042710} (\bibinfo {year} {2001})}\BibitemShut {NoStop}%
\bibitem [{\citenamefont {Talu}\ and\ \citenamefont
  {Myers}(2001)}]{TALU200183}%
  \BibitemOpen
  \bibfield  {author} {\bibinfo {author} {\bibfnamefont {O.}~\bibnamefont
  {Talu}}\ and\ \bibinfo {author} {\bibfnamefont {A.~L.}\ \bibnamefont
  {Myers}},\ }\bibfield  {title} {\bibinfo {title} {Reference potentials for
  adsorption of helium, argon, methane, and krypton in high-silica zeolites},\
  }\href {https://doi.org/https://doi.org/10.1016/S0927-7757(01)00628-8}
  {\bibfield  {journal} {\bibinfo  {journal} {Colloids and Surfaces A:
  Physicochemical and Engineering Aspects}\ }\textbf {\bibinfo {volume}
  {187-188}},\ \bibinfo {pages} {83 } (\bibinfo {year} {2001})}\BibitemShut
  {NoStop}%
\bibitem [{\citenamefont {Pitz}\ \emph {et~al.}(2009)\citenamefont {Pitz},
  \citenamefont {Wertepny},\ and\ \citenamefont {Perram}}]{PhysRevA.80.062718}%
  \BibitemOpen
  \bibfield  {author} {\bibinfo {author} {\bibfnamefont {G.~A.}\ \bibnamefont
  {Pitz}}, \bibinfo {author} {\bibfnamefont {D.~E.}\ \bibnamefont {Wertepny}},\
  and\ \bibinfo {author} {\bibfnamefont {G.~P.}\ \bibnamefont {Perram}},\
  }\bibfield  {title} {\bibinfo {title} {Pressure broadening and shift of the
  cesium ${D}_{1}$ transition by the noble gases and {N}$_{2}$, {H}$_{2}$,
  {H}{D}, {D}$_{2}$, {C}{H}$_{4}$, {C}$_{2}${H}${}_{6}$, {C}{F}${}_{4}$, and
  $^{3}\text{H}\text{e}$},\ }\href {https://doi.org/10.1103/PhysRevA.80.062718}
  {\bibfield  {journal} {\bibinfo  {journal} {Phys. Rev. A}\ }\textbf {\bibinfo
  {volume} {80}},\ \bibinfo {pages} {062718} (\bibinfo {year}
  {2009})}\BibitemShut {NoStop}%
\bibitem [{\citenamefont {Pitz}\ \emph {et~al.}(2010)\citenamefont {Pitz},
  \citenamefont {Fox},\ and\ \citenamefont {Perram}}]{PhysRevA.82.042502}%
  \BibitemOpen
  \bibfield  {author} {\bibinfo {author} {\bibfnamefont {G.~A.}\ \bibnamefont
  {Pitz}}, \bibinfo {author} {\bibfnamefont {C.~D.}\ \bibnamefont {Fox}},\ and\
  \bibinfo {author} {\bibfnamefont {G.~P.}\ \bibnamefont {Perram}},\ }\bibfield
   {title} {\bibinfo {title} {Pressure broadening and shift of the cesium
  ${D}_{2}$ transition by the noble gases and {N}$_{2}$, {H}$_{2}$, {H}{D},
  {D}$_{2}$, {C}{H}$_{4}$, {C}$_{2}${H}${}_{6}$, {C}{F}${}_{4}$, and
  $^{3}\mathrm{He}$ with comparison to the ${D}_{1}$ transition},\ }\href
  {https://doi.org/10.1103/PhysRevA.82.042502} {\bibfield  {journal} {\bibinfo
  {journal} {Phys. Rev. A}\ }\textbf {\bibinfo {volume} {82}},\ \bibinfo
  {pages} {042502} (\bibinfo {year} {2010})}\BibitemShut {NoStop}%
\bibitem [{\citenamefont {{Margrave}}(1964)}]{1964JChEd..41R.590M}%
  \BibitemOpen
  \bibfield  {author} {\bibinfo {author} {\bibfnamefont {J.~L.}\ \bibnamefont
  {{Margrave}}},\ }\bibfield  {title} {\bibinfo {title} {{Vapour pressure of
  the chemical elements (Nesmeyanov, A. N.)}},\ }\href
  {https://doi.org/10.1021/ed041pA590.3} {\bibfield  {journal} {\bibinfo
  {journal} {Journal of Chemical Education}\ }\textbf {\bibinfo {volume}
  {41}},\ \bibinfo {pages} {A590} (\bibinfo {year} {1964})}\BibitemShut
  {NoStop}%
\bibitem [{\citenamefont {Mitroy}\ and\ \citenamefont
  {Tang}(2013)}]{PhysRevA.88.052515}%
  \BibitemOpen
  \bibfield  {author} {\bibinfo {author} {\bibfnamefont {J.}~\bibnamefont
  {Mitroy}}\ and\ \bibinfo {author} {\bibfnamefont {L.-Y.}\ \bibnamefont
  {Tang}},\ }\bibfield  {title} {\bibinfo {title} {Tune-out wavelengths for
  metastable helium},\ }\href {https://doi.org/10.1103/PhysRevA.88.052515}
  {\bibfield  {journal} {\bibinfo  {journal} {Phys. Rev. A}\ }\textbf {\bibinfo
  {volume} {88}},\ \bibinfo {pages} {052515} (\bibinfo {year}
  {2013})}\BibitemShut {NoStop}%
\bibitem [{\citenamefont {Astapenko}(2013)}]{astapenko2013polarization}%
  \BibitemOpen
  \bibfield  {author} {\bibinfo {author} {\bibfnamefont {V.}~\bibnamefont
  {Astapenko}},\ }\href@noop {} {\emph {\bibinfo {title} {Polarization
  Bremsstrahlung on Atoms, Plasmas, Nanostructures and Solids}}},\
  Vol.~\bibinfo {volume} {72}\ (\bibinfo  {publisher} {Springer Science \&
  Business Media},\ \bibinfo {year} {2013})\BibitemShut {NoStop}%
\bibitem [{\citenamefont {Kien}\ \emph {et~al.}(2013)\citenamefont {Kien},
  \citenamefont {Schneeweiss},\ and\ \citenamefont
  {Rauschenbeutel}}]{LeKien2013}%
  \BibitemOpen
  \bibfield  {author} {\bibinfo {author} {\bibfnamefont {F.~L.}\ \bibnamefont
  {Kien}}, \bibinfo {author} {\bibfnamefont {P.}~\bibnamefont {Schneeweiss}},\
  and\ \bibinfo {author} {\bibfnamefont {A.}~\bibnamefont {Rauschenbeutel}},\
  }\bibfield  {title} {\bibinfo {title} {Dynamical polarizability of atoms in
  arbitrary light fields: general theory and application to cesium},\ }\href
  {https://doi.org/10.1140/epjd/e2013-30729-x} {\bibfield  {journal} {\bibinfo
  {journal} {The European Physical Journal D}\ }\textbf {\bibinfo {volume}
  {67}},\ \bibinfo {pages} {92} (\bibinfo {year} {2013})}\BibitemShut {NoStop}%
\bibitem [{\citenamefont {Stalnaker}\ \emph {et~al.}(2006)\citenamefont
  {Stalnaker}, \citenamefont {Budker}, \citenamefont {Freedman}, \citenamefont
  {Guzman}, \citenamefont {Rochester},\ and\ \citenamefont
  {Yashchuk}}]{PhysRevA.73.043416}%
  \BibitemOpen
  \bibfield  {author} {\bibinfo {author} {\bibfnamefont {J.~E.}\ \bibnamefont
  {Stalnaker}}, \bibinfo {author} {\bibfnamefont {D.}~\bibnamefont {Budker}},
  \bibinfo {author} {\bibfnamefont {S.~J.}\ \bibnamefont {Freedman}}, \bibinfo
  {author} {\bibfnamefont {J.~S.}\ \bibnamefont {Guzman}}, \bibinfo {author}
  {\bibfnamefont {S.~M.}\ \bibnamefont {Rochester}},\ and\ \bibinfo {author}
  {\bibfnamefont {V.~V.}\ \bibnamefont {Yashchuk}},\ }\bibfield  {title}
  {\bibinfo {title} {Dynamic stark effect and forbidden-transition spectral
  line shapes},\ }\href {https://doi.org/10.1103/PhysRevA.73.043416} {\bibfield
   {journal} {\bibinfo  {journal} {Phys. Rev. A}\ }\textbf {\bibinfo {volume}
  {73}},\ \bibinfo {pages} {043416} (\bibinfo {year} {2006})}\BibitemShut
  {NoStop}%
\bibitem [{\citenamefont {Cohen-Tannoudji}(1996)}]{CohenTannoudji1996}%
  \BibitemOpen
  \bibfield  {author} {\bibinfo {author} {\bibfnamefont {C.~N.}\ \bibnamefont
  {Cohen-Tannoudji}},\ }\bibfield  {title} {\bibinfo {title} {The autler-townes
  effect revisited},\ }in\ \href {https://doi.org/10.1007/978-1-4612-2378-8_11}
  {\emph {\bibinfo {booktitle} {Amazing Light}}}\ (\bibinfo  {publisher}
  {Springer New York},\ \bibinfo {year} {1996})\ pp.\ \bibinfo {pages}
  {109--123}\BibitemShut {NoStop}%
\bibitem [{\citenamefont {Fendel}\ \emph {et~al.}(2007)\citenamefont {Fendel},
  \citenamefont {Bergeson}, \citenamefont {Udem},\ and\ \citenamefont
  {H\"{a}nsch}}]{Fendel:07}%
  \BibitemOpen
  \bibfield  {author} {\bibinfo {author} {\bibfnamefont {P.}~\bibnamefont
  {Fendel}}, \bibinfo {author} {\bibfnamefont {S.~D.}\ \bibnamefont
  {Bergeson}}, \bibinfo {author} {\bibfnamefont {T.}~\bibnamefont {Udem}},\
  and\ \bibinfo {author} {\bibfnamefont {T.~W.}\ \bibnamefont {H\"{a}nsch}},\
  }\bibfield  {title} {\bibinfo {title} {Two-photon frequency comb spectroscopy
  of the 6s-8s transition in cesium},\ }\href
  {https://doi.org/10.1364/OL.32.000701} {\bibfield  {journal} {\bibinfo
  {journal} {Opt. Lett.}\ }\textbf {\bibinfo {volume} {32}},\ \bibinfo {pages}
  {701} (\bibinfo {year} {2007})}\BibitemShut {NoStop}%
\bibitem [{\citenamefont {Wu}\ \emph {et~al.}(2013)\citenamefont {Wu},
  \citenamefont {Liu}, \citenamefont {Wu}, \citenamefont {Lee},\ and\
  \citenamefont {Cheng}}]{Wu:13}%
  \BibitemOpen
  \bibfield  {author} {\bibinfo {author} {\bibfnamefont {C.-M.}\ \bibnamefont
  {Wu}}, \bibinfo {author} {\bibfnamefont {T.-W.}\ \bibnamefont {Liu}},
  \bibinfo {author} {\bibfnamefont {M.-H.}\ \bibnamefont {Wu}}, \bibinfo
  {author} {\bibfnamefont {R.-K.}\ \bibnamefont {Lee}},\ and\ \bibinfo {author}
  {\bibfnamefont {W.-Y.}\ \bibnamefont {Cheng}},\ }\bibfield  {title} {\bibinfo
  {title} {Absolute frequency of cesium 6{S}-8{S} 822 nm two-photon transition
  by a high-resolution scheme},\ }\href {https://doi.org/10.1364/OL.38.003186}
  {\bibfield  {journal} {\bibinfo  {journal} {Opt. Lett.}\ }\textbf {\bibinfo
  {volume} {38}},\ \bibinfo {pages} {3186} (\bibinfo {year}
  {2013})}\BibitemShut {NoStop}%
\bibitem [{\citenamefont {Kramida}\ \emph {et~al.}(2019)\citenamefont
  {Kramida}, \citenamefont {Ralchenko}, \citenamefont {Reader},\ and\
  \citenamefont {{NIST ASD Team}}}]{NISTASD}%
  \BibitemOpen
  \bibfield  {author} {\bibinfo {author} {\bibfnamefont {A.}~\bibnamefont
  {Kramida}}, \bibinfo {author} {\bibfnamefont {Y.}~\bibnamefont {Ralchenko}},
  \bibinfo {author} {\bibfnamefont {J.}~\bibnamefont {Reader}},\ and\ \bibinfo
  {author} {\bibnamefont {{NIST ASD Team}}},\ }\href
  {https://physics.nist.gov/asd} {} (\bibinfo {year} {2019}),\ \bibinfo {note}
  {{NIST Atomic Spectra Database (ver. 5.7.1), [Online]. Available:
  {\tt{https://physics.nist.gov/asd}} [2019, November 10]. National Institute
  of Standards and Technology, Gaithersburg, MD.}}\BibitemShut {Stop}%
\bibitem [{\citenamefont {Huber}\ and\ \citenamefont
  {Sandeman}(1986)}]{Huber1986}%
  \BibitemOpen
  \bibfield  {author} {\bibinfo {author} {\bibfnamefont {M.~C.~E.}\
  \bibnamefont {Huber}}\ and\ \bibinfo {author} {\bibfnamefont {R.~J.}\
  \bibnamefont {Sandeman}},\ }\bibfield  {title} {\bibinfo {title} {The
  measurement of oscillator strengths},\ }\href
  {https://doi.org/10.1088/0034-4885/49/4/002} {\bibfield  {journal} {\bibinfo
  {journal} {Reports on Progress in Physics}\ }\textbf {\bibinfo {volume}
  {49}},\ \bibinfo {pages} {397} (\bibinfo {year} {1986})}\BibitemShut
  {NoStop}%
\bibitem [{\citenamefont {Ligare}(2010)}]{doi:10.1119/1.3417868}%
  \BibitemOpen
  \bibfield  {author} {\bibinfo {author} {\bibfnamefont {M.}~\bibnamefont
  {Ligare}},\ }\bibfield  {title} {\bibinfo {title} {Classical thermodynamics
  of particles in harmonic traps},\ }\href {https://doi.org/10.1119/1.3417868}
  {\bibfield  {journal} {\bibinfo  {journal} {American Journal of Physics}\
  }\textbf {\bibinfo {volume} {78}},\ \bibinfo {pages} {815} (\bibinfo {year}
  {2010})}\BibitemShut {NoStop}%
\bibitem [{\citenamefont {Krainov}\ and\ \citenamefont
  {Smirnov}(2019)}]{Krainov2019}%
  \BibitemOpen
  \bibfield  {author} {\bibinfo {author} {\bibfnamefont {V.}~\bibnamefont
  {Krainov}}\ and\ \bibinfo {author} {\bibfnamefont {B.~M.}\ \bibnamefont
  {Smirnov}},\ }\bibinfo {title} {Atomic and molecular radiative processes}\
  (\bibinfo  {publisher} {Springer International Publishing},\ \bibinfo {year}
  {2019})\ Chap.\ \bibinfo {chapter} {1.2.4}, pp.\ \bibinfo {pages}
  {26--30}\BibitemShut {NoStop}%
\bibitem [{\citenamefont {Metcalf}\ and\ \citenamefont {Van~der
  Straten}(1999)}]{metcalf1999laser}%
  \BibitemOpen
  \bibfield  {author} {\bibinfo {author} {\bibfnamefont {H.~J.}\ \bibnamefont
  {Metcalf}}\ and\ \bibinfo {author} {\bibfnamefont {P.}~\bibnamefont {Van~der
  Straten}},\ }\bibinfo {title} {Laser cooling and trapping}\ (\bibinfo
  {publisher} {Springer-Verlag},\ \bibinfo {address} {New York},\ \bibinfo
  {year} {1999})\ Chap.~\bibinfo {chapter} {4}, pp.\ \bibinfo {pages}
  {53--56}\BibitemShut {NoStop}%
\end{thebibliography}%


\onecolumngrid
\section*{Supporting Online Material}
\appendix
\section{Experimental Details}
\label{sec:exp-details}
\begin{figure}[b]
    \centering
    \includegraphics[width=0.6\textwidth]{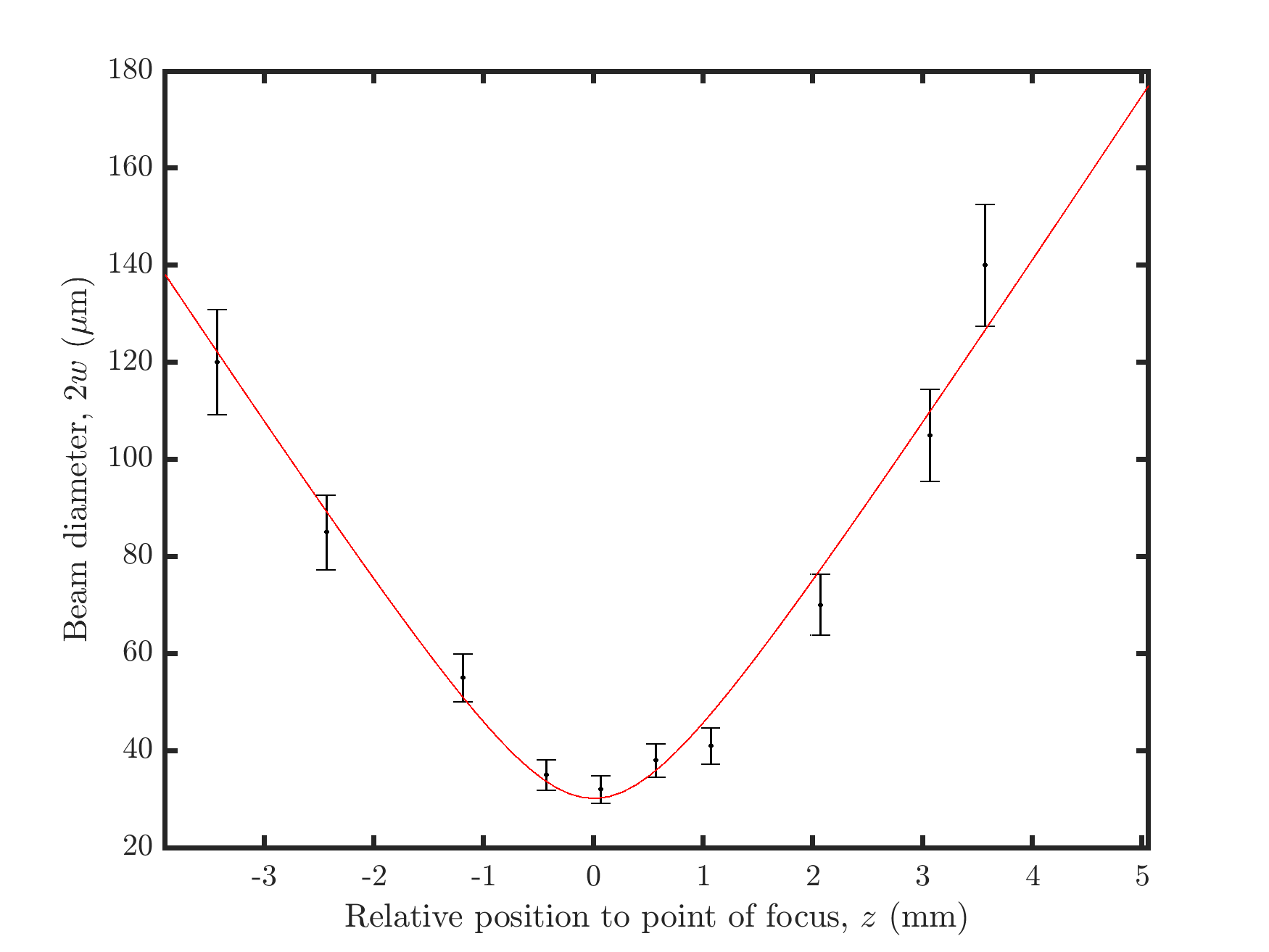}
    \caption{Beam diameter (twice the beam width \(w\)) of the probe beam over relative distance along the axis of the probe beam. A fit to the spot size of the form \(w(z) = w_0 \sqrt{1+(z/z_R)^2}\) is shown in red, where \(w_0=16\)~\(\mu m\) is the beam waist, and \(z_R=91\)~\(\mu m\) is the Rayleigh range.}
    \label{fig:BEAM_WAIST}
\end{figure}

\begin{figure}
    \centering
    \includegraphics[width=0.7\textwidth]{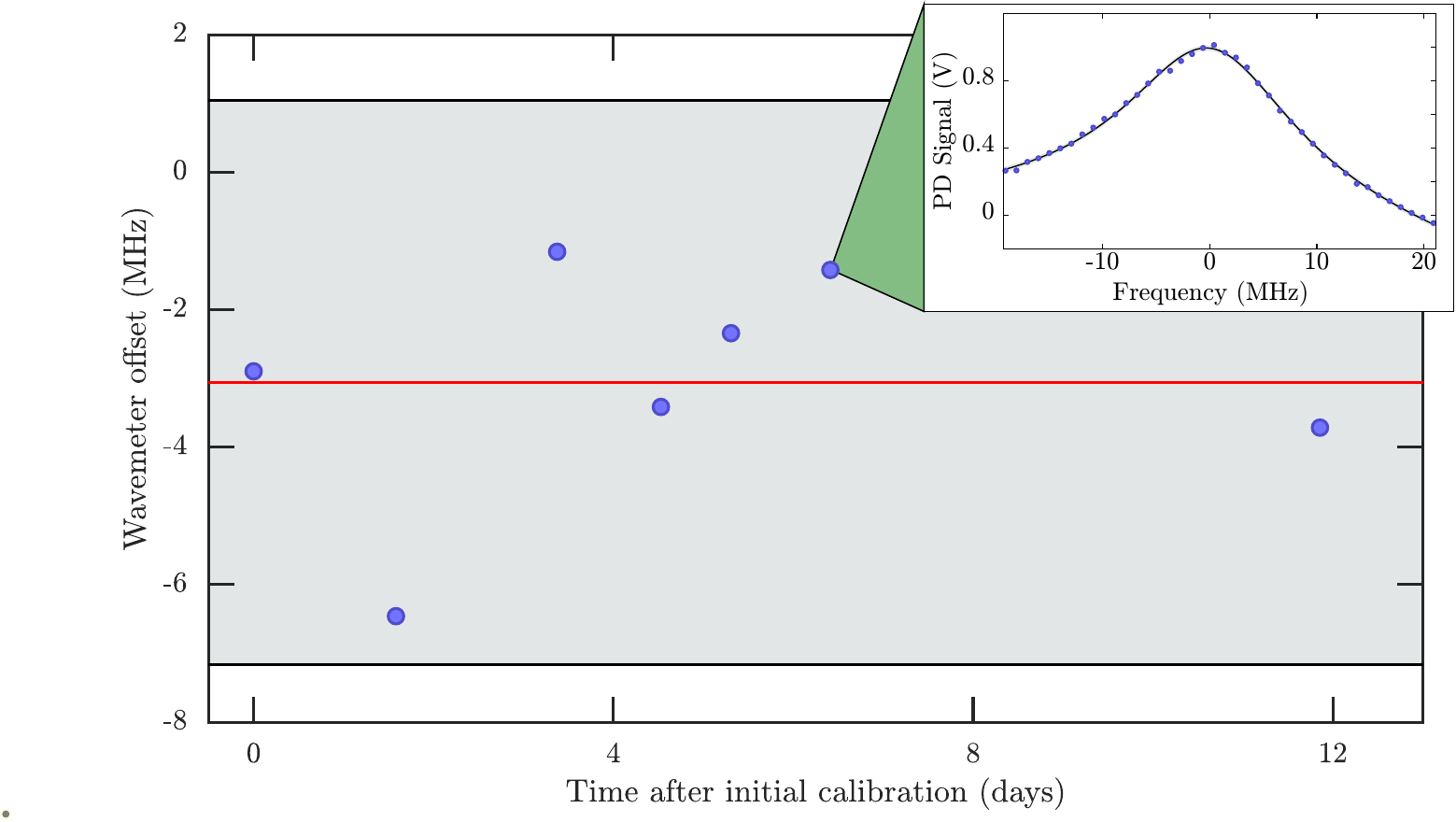}
    \caption{Measured wavemeter offset versus time. Blue circles mark empirical data, with the red center line showing the mean of the data \(-3.01\)~MHz, and the black lines and shaded region indicating an uncertainty of \(4.1\)~MHz, which arises from the specified manufacture error and the distribution of the data. Standard deviation of the points is 1.7~MHz. Inset shows a single measurement using a fit of a Lorentzian with a second order polynomial background model to the cesium saturated absorption spectroscopy  of the \(6^2S_{\nicefrac{1}{2}}, F=4\) to the \(6^2P_{\nicefrac{3}{2}}, F=4\) and \(6^2P_{\nicefrac{3}{2}}, F=5\)  crossover line, with the measured frequency (from the wavemeter) offset by \(351,721,835.0\)~MHz, the measured frequency of Tanner and Wieman \cite{pmid9900545}.}
    \label{fig:wm_model}
\end{figure}

\textbf{Probe Beam:} The probe beam consists of a Gaussian beam focused down, over a $\sim$700~mm distance, to a spot with a \(16\)~\(\mu\)m beam waist and Rayleigh range of \(91\)~\(\mu\)m, relative to the mean width of the atoms \(8\)~\(\mu\)m. See Fig.~\ref{fig:BEAM_WAIST} for beam profile along the axis of the probe beam, along with a fit to the spot size of the form \(w(z) = w_0 \sqrt{1+(z/z_R)^2}\) is shown in red, where \(w_0=16\)~\(\mu\)m is the beam waist, and \(z_R=91\)~\(\mu\)m is the Rayleigh range. The probe beam power averages \(31\)~mW with a standard deviation of \(4\)~mW, which gives an intensity at the focus of \(3.86 \times 10^7\)~W/m\(^2\). The atoms are exposed to the beam for a 20~\(s\) period.

\textbf{Apparatus:} The laser used was an M squared SolsTiS, a widely tuneable (700 to 1000 nm) Ti:Sapphire laser, with an M squared ECD-X doubling cavity (providing light in a 350 to 500 nm range). The laser was stabilised at a specific frequency via a proportional-integral-derivative (PID) feedback loop between a HighFinesse/\r{A}ngstrom WS8-2 wavemeter, measuring the laser's output, and the lab computer which controlled the scanning cavity within the SolsTiS. The power of the probe beam was measured just before the experimental chamber over time via the use of a photodiode, in order to calibrate the final signal with respect to incident power. 


\textbf{Calibration of the Wavemeter:} The wavemeter was periodically calibrated by using saturated absorption spectroscopy in a cell with the non-doubled light (700 to 1000~nm) to perform  measurement of a known cesium crossover transition from \(6^2S_{\nicefrac{1}{2}}, F=4\) to between \(6^2P_{\nicefrac{3}{2}}, F=4\) and \(6^2P_{\nicefrac{3}{2}}, F=5\). This reference transition was constrained to be \(351,721,835.0(1)\)~MHz, equivalently \(852.3566870(3)\)~nm, by the measurement of Tanner and Wieman \cite{pmid9900545}. The reason this particular transition was chosen as a reference was that after applying the doubling cavity the light from this transition is \(\sim\)1.5~nm below the theoretically expected and experimentally measured wavelength of the transition of interest. From the manufacturer specifications we know that a calibration within \(2.0\)~nm gives the wavemeter an accuracy of \(2.0\)~MHz \cite{wstechnical}, however, as we are calibrating before the doubling cavity we must double this error to \(4.0\)~MHz.
The calibration shift in the wavemeter over time is shown in Fig.~\ref{fig:wm_model}. We correct for systematic wavemeter drift using the average of the calibrations, a shift of \(-3.01\)~MHz. The standard error in the distribution of the measurements was added to the fundamental wavemeter uncertainty to obtain a final uncertainty in this shift of \(4.1\)~MHz.  

\textbf{Detection:} The atoms are imaged in the far-field with full three dimensional resolution using an \(80\)~mm diameter micro-channel plate and delay line detector located \(\sim\)\(850\)~mm below trap centre, with a spatial resolution of \(\sim\)120~\(\mu m\) and temporal resolution of \(\sim\)3~\(\mu s\) \cite{Manning:10}.
 
 \textbf{Trap:} We use a magnetic bi-planar quadrupole Ioffe trap \cite{Dall2007}, which has trapping frequencies \(\omega_{(x,y,z)}/2\pi=\big(53.5(6),426.56(5),430.30(6)\big)\)~Hz. The trap frequencies of the magnetic trap were determined by inducing oscillations in the trap and then outcoupling portions of the atoms with RF pulses over time. From the measured 3D oscillations of the atoms we extract the trap frequencies.
 
 
\textbf{Alignment of the Probe Beam:} We also use our trap frequency measurement technique as a tool to align the probe beam. We looked for a difference in the trap frequency between when the probe beam was and was not applied to the atoms. This frequency change is due to the dynamic polarizability the probe beam exerts on the atoms, adding an additional trapping potential that is approximately harmonic, which in turn purely depends upon the wavelength and the intensity of the light applied. Hence by keeping the wavelength fixed we were able to maximize the alignment on the atoms via maximizing the change in trap frequency.

\textbf{Fundamental light:} To ensure we are not observing a two photon process caused by the non-doubled fundamental light of the laser we use a range of measures: there is a filter which blocks wavelengths around the fundamental range just before the optic fiber that directs the light to the experimental chamber; the laser table and experimental chamber are completely isolated from each other; and we observe linear scaling of the signal amplitude with power in agreement with a single photon process.

\textbf{Radio Frequency Outcoupling:} The pulsed atom laser used to outcouple the atoms consists of a series of radio frequency  approximately 20~\(\mu\)s in length corresponding to a Fourier width of \(\sim 300\)~kHz \cite{Manning:10} which is much larger than the frequency width of the atomic distribution in the trap, which is given by the thermal energy distribution to be approximately \(10\)~kHz.


\section{Systematic Frequency Shifts}
\label{sec:freq_shifts}
\begin{figure}
    \centering
    \includegraphics[width=0.6\textwidth]{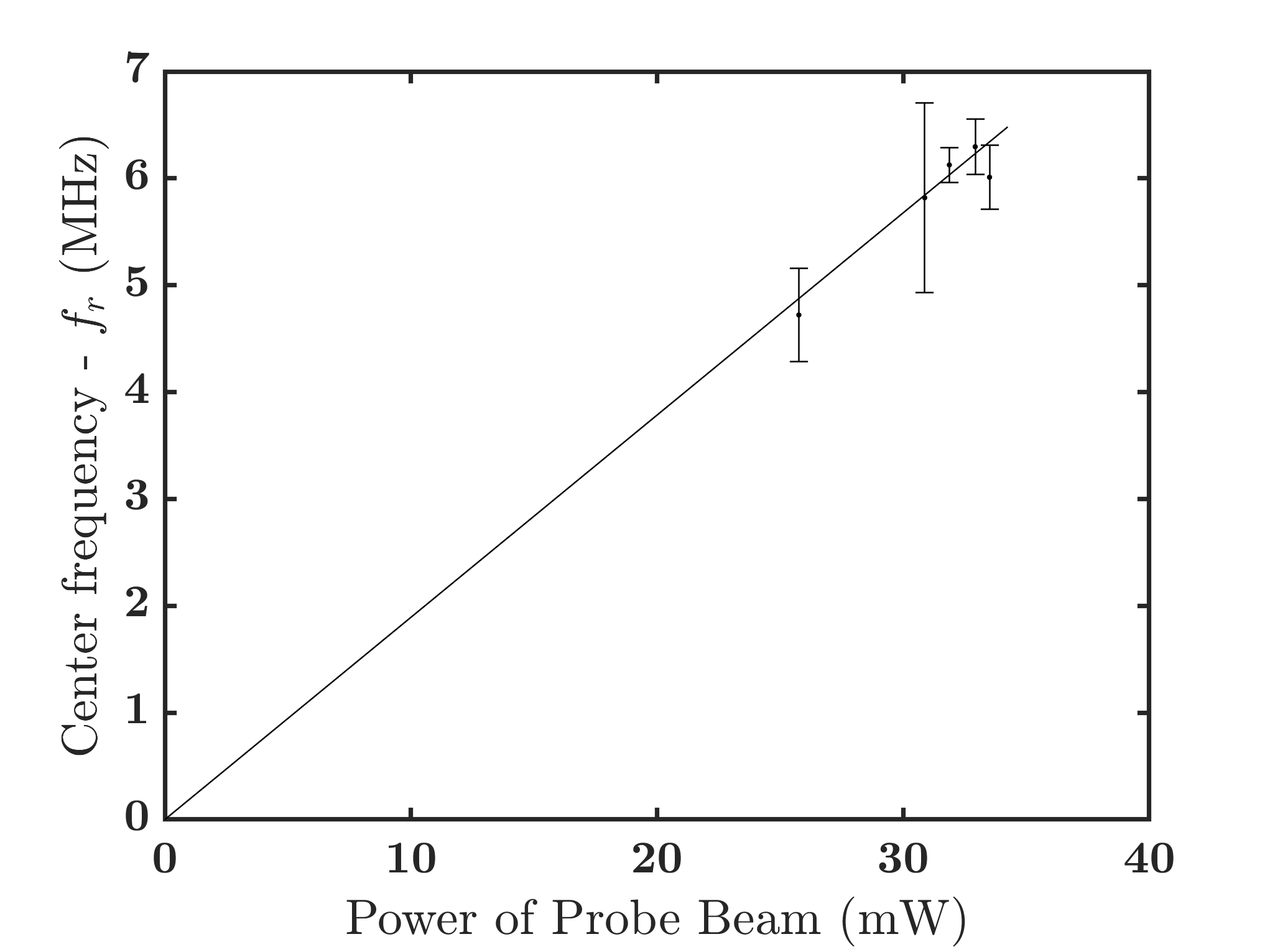}
    \caption{Frequency center of the distribution relative to field free value \(f_r=700,939,271\)~MHz as a function of applied probe beam power. Note the reason the uncertainty of the data in this figure is less then implied by Fig.~\ref{fig:wm_model} is due to the shorter time scale these data points where taken over and that the systematic uncertainty in the wavemeter, which is the major source of uncertainty in Fig.~\ref{fig:wm_model}, does not effect the relative uncertainty of these points. See Sec.~\ref{sec:laser_linewidth} for more detail.}
    \label{fig:ac_stark} %
\end{figure}
         \textbf{Zeeman shift:}
         The transition frequency is shifted by the Zeeman effect due to the energies of the initial and final states being shifted by different amounts due to external magnetic fields. The magnitude of the external field at the atoms is measured to be approximately 0.613(1)~G, by using a swept RF pulse to measure the Zeeman splitting between the \(\MetastableState, \,m_J=+1\) and \(m_J=0\) states. Note the magnitude of the magnetic field does not change between the direct detection and heating method. At field strengths of this size only the linear Zeeman effects will be relevant to our experimental precision. The frequency shift due to the linear Zeeman effect has the functional form 
         \begin{align}
             \Delta f &= \frac{\mu_B}{h} B (m_{J}^{e} g_{j}^{e} - m_{J}^{g} g_{j}^{g}),
         \end{align}
         where \(\mu_B\) is the Bohr magneton, \(h\) is the Planck constant, \(B\) is the magnetic field magnitude, \(g_{j}^{e}\) and \(g_{j}^{g}\) are the Land\'e \(g\)-factors of the excited and ground states respectively, and \(m_J^{e}\) and \(m_J^{g}\) are their respective magnetic quantum numbers. The atoms are initially in the \(\MetastableState, \, m_{J}^{g}=+1\) state with \(g_{j}^{g} = 2.0\) and are excited to the \(\UpperState, \, m_{J}^{e}=0\) state with \(g_{j}^{e} = 2.0\). Hence the Zeeman shift for our transition is \(\Delta f_{Zeeman} = - 1.715(3)\)~MHz.\\
         
         \textbf{AC Stark shift:}
         For the tested experimental range the AC Stark shift is linearly proportional to the intensity of the probe beam at the atoms. Note we do not use any other light source in our trapping potential. As the focus of the probe beam is kept constant, the AC stark shift is linearly dependant on the total power in the beam. The power of the probe beam was measured using a calibrated photodiode. The frequency of the transitions was measured at a range of probe beam powers, as shown in Fig.~\ref{fig:ac_stark}. The data can then be used to linearly extrapolate to a field-free value of the frequency, with an uncertainty determined by the confidence interval of the fit. For the direct detection method we find a shift of \(\Delta f_{AC,Probe} = 8.6(1.5)\)~MHz and for the heating method \(\Delta f_{AC,Probe} = 5.9 (1.6)\)~MHz. The reason these values differ is the applied probe beam's focus size varied between the methods.
         
         \textbf{DC Stark shift:} For a multi-electron atom in either ground or low excited state in the presence of a weak static electric field its energy levels will be shifted as \(\Delta E = -\frac{1}{2} \alpha_s \mathbf{E}_{dc}^2\), where \(\alpha_s\) is the static polarisability of the atoms (in their current state) and \(\mathbf{E}_{dc}\) is the dc electric field of strength. As the electric field is kept constant the shift in frequency due to this effect is given by \(\Delta f_{DC} = -\frac{1}{2h} \Delta \alpha_s \mathbf{E}_{dc}^2\). For our case we can constrain from direct measurement that any static electric field in our experimental chamber has \(\mathbf{E}_{dc}<2\)~kV/m and \(\Delta \alpha < 6\times 10^{-50}\)~C\(^3\)m\(^3\)/J\(^2\) \cite{Kondratjev2010}, thus \(\Delta f_{DC}<10^{-10}\)~Hz. \\
         
         \textbf{Mean field shift:} The interactions between atoms in a BEC can also shift spectral lines, termed the mean field shift. For bosonic particles with sufficiently low temperatures such that only \(s\)-wave scattering will occur, which for the case of He\(^*\) corresponds to temperatures less than \(100\)~mK \cite{Julienne:89}, the density dependent mean field shift of the atomic energy level in a degenerate homogeneous system is given by \cite{PhysRevLett.81.3807}
         \begin{align}
             \Delta E = \frac{8 \pi \hbar^2 a n}{m},\label{eqn:mean_field_shift}
         \end{align}
         where \(a\) is the scattering length of the atoms in their current state, \(m\) is the mass of the atoms and \(n\) is the density of the atoms. Note that Eqn.~\ref{eqn:mean_field_shift} assumes only the elastic contribution to the collisional shift is relevant, which is valid for weak excitations \cite{PhysRevLett.81.3807}. The frequency shift induced in the spectroscopic transitions, neglecting inelastic processes, is hence
         \begin{align}
             \Delta f_{mf} = \frac{4 \hbar}{m} \left(n_f a_{f-f} + n_i a_{i-f} - n_f a_{i-f} - n_i a_{i-i}\right),
         \end{align} 
         where \(n_i\) and \(n_f\) are the density of the atoms in the initial and final states respectively and \(a_{i-i}\), \(a_{i-f}\) and \(a_{f-f}\) are the scattering lengths between the initial and initial, initial and final, and final and final states respectively. As the transition we are considering is extremely weak the density of the final state would be negligible, and thus \(\Delta f_{mf} \approx  \frac{4 \hbar n_i}{m} (a_{i-f} - a_{i-i})\). The scattering length of the \(\MetastableState-\MetastableState\) state collisions is of the order of \(10\)~nm \cite{PhysRevLett.96.023203,PhysRevLett.93.090409}, while the scattering length of the \(\MetastableState-\UpperState\) collisions can be calculated to be of order \(1\)~nm \cite{PhysRevA.48.546,PhysRevA.64.042710,TALU200183}. The average density of the atoms can be calculated from the atom number and the trap frequencies to be on the order of \(10^{19}\)~m\(^{-3}\). This gives us a mean field shift of the order \(\Delta f_{mf} \sim -10\)~kHz.\\
         
\begin{figure}[b]
    \centering
    \includegraphics[width=0.6\textwidth]{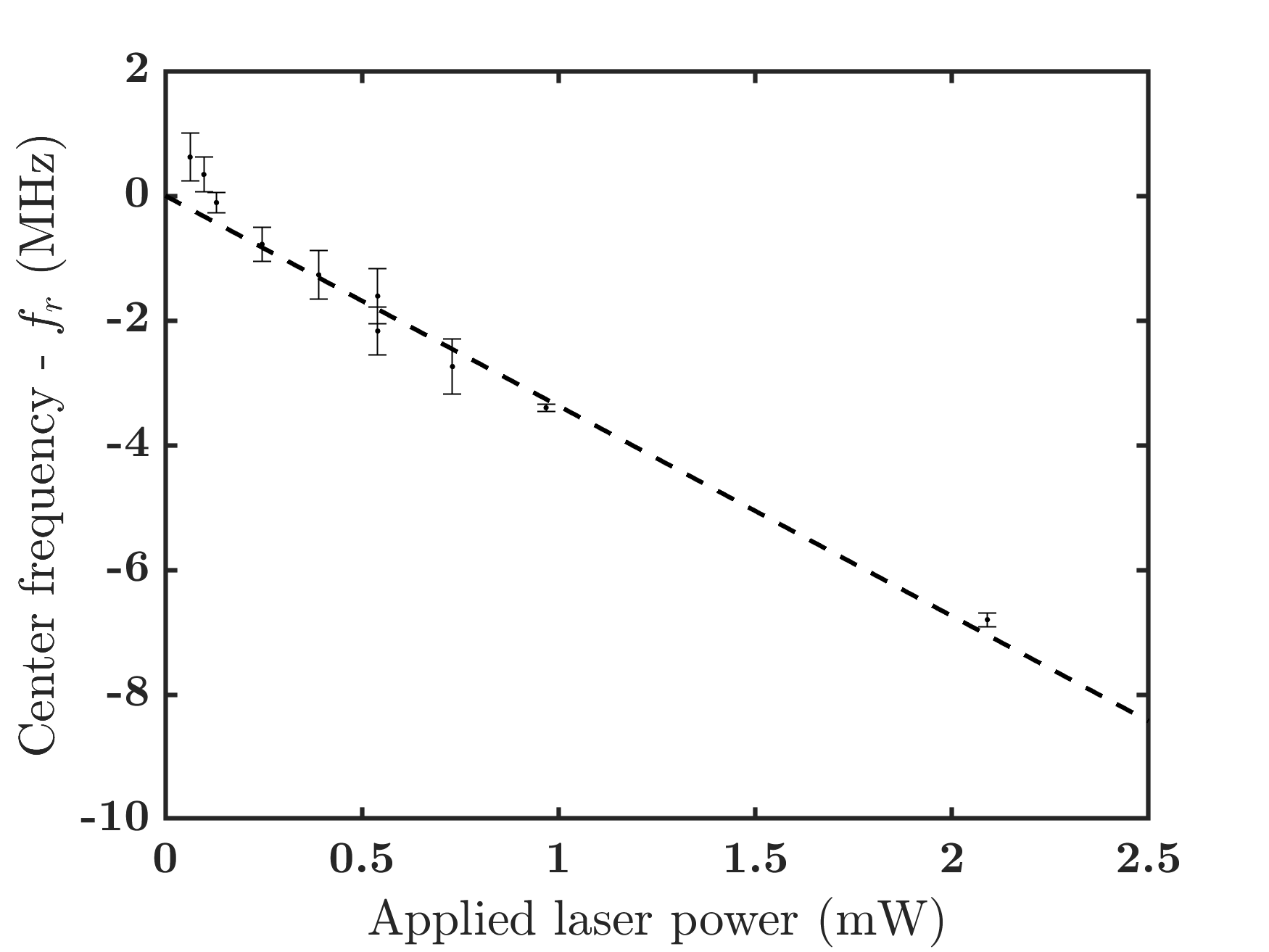}
    \caption{Measured center frequency of cesium calibration transition relative to the extrapolated theory free value \(f_r\) as a function of applied laser power. The dashed line represents linear best fit, with equation \(f_{rel} = -3.37\times 10^{-3} P\) where \(f_{rel}\) is the relative frequency in MHz and \(P\) is the applied power in \(\mu\)W. Note that like Fig.~\ref{fig:ac_stark} the uncertainty in these points are less than implied by Fig.~\ref{fig:wm_model} due to the different time scales they were taken over and as the systematic uncertainty does not affect the relative uncertainty of these points. See Sec.~\ref{sec:laser_linewidth} for more detail.}
    \label{fig:cs_stark_shift}
\end{figure}

         \textbf{Recoil shift:}
         Due to the conservation of momentum, when a photon is absorbed during an atomic transition the photon's momentum is imparted onto the atom. The momentum of the photon, and thus the change in momentum of the atom, is given by \(\Delta p = \frac{hf}{c}\), where \(f\) is the frequency of the absorbed photon and \(c\) is the speed of light. This increase in the kinetic energy of the atom must come from the photon, implying that there must be a shift in the energy of the photon in order to compensate for this imparted energy. The recoil shift of the photon's energy is \(\Delta E = \frac{1}{2m} \left( \frac{hf}{c} \right)^2\), where \(m\) is the atomic mass. The frequency of the transition was measured to be \(700,939,271(5)\)~MHz giving a recoil shift of \(0.273\)~MHz. As the relative uncertainty is on the order of parts per hundred-million the uncertainty within the recoil shift is well below \(1\)~kHz.\\

         \textbf{Cesium Cell offset:}  There are two main systematic shifts which occur in the cesium cell used to spectroscopically reference the wavemeter and laser: the AC Stark shift due to the probe laser and the vapour (or pressure) shift due to collisions within the cell. The AC Stark shift can be determined in the same manner as it was for the probe beam, by varying the power and measuring the change in the center frequency and then extrapolating to a theory free value, see Fig.~\ref{fig:cs_stark_shift}. The normal applied laser power is \(560\)~\(\mu\)W which gives an AC stark shift of \(\Delta f_{AC,Cs} = -1.9(4)\)~MHz. The pressure shift in the cell can be constrained using literature values, which state that the pressure shift is less than \(30\)~MHz/torr \cite{PhysRevA.80.062718,PhysRevA.82.042502}. The cell was at a temperature of \(84(1)\)~\(^\circ C\) which corresponds to a vapour pressure of \(2.00(2)\times 10^{-4}\)~torr \cite{1964JChEd..41R.590M}. Thus the vapour pressure shift in the cell is constrained to be \(\Delta f_{pressure} < 6 \times 10^{-3}\)~MHz.

\section{Polarisability effects on lineshape}
It can be see in Fig.~3 of the main text that the signal decays to a non-zero negative value, with a p-value of \(8.4 \times 10^{-7}\). This is most likely due to the repulsive dipole lensing caused by the probe laser, which leads to a slightly smaller proportion of atoms being detected.

We note that this may seem to imply that the dynamic polarisabilty (or ac-Stark shift) could effect the measured line shape. However, we believe this is not the case as the dynamic polarisability of the \(\MetastableState\) is approximately constant over the transition \cite{PhysRevA.88.052515,astapenko2013polarization,LeKien2013}, and hence so to is the approximate downward shift of the signal due to the dipole lensing. Furthermore the downward shift is small in comparison to the signal amplitude, the negative offset has a value of about \(2\%\) of the maximum signal. 

To see intuitively why the dynamic polarisability is approximately constant over the transitions linewidth consider that the dynamic polarisability of an atom \(\alpha(\omega)\), at a given frequency \(\omega\) and in a particular state, is give by the equation \cite{astapenko2013polarization,LeKien2013}
\begin{align}
    \alpha (\omega) &= \frac{e^2}{m_e} \sum_n \frac{f_n}{\omega_n^2-\omega^2- i \omega \delta_n}
\end{align}
where we are summing over all possible electronic transitions, \(e\) and \(m_e\) are the charge and mass of an electron, \(\delta_n\), \(f_n\) and \(\omega_n\) are the linewidth, oscillator strength, and center frequency of each transition. From this we can see that the main contributing factor to a particular transitions effect on the polarisability, aside from its detuning is its oscillator strength.

The oscillator strengths \(\MetastableState \rightarrow 3^{3\!}P\) and \(\MetastableState \rightarrow 2^{3\!}P\) are approximately \(12\) orders of mangitude greater than that of the \(\MetastableState \rightarrow \UpperState\) transition.  The \(\MetastableState \rightarrow 3^{3\!}P\) and \(\MetastableState \rightarrow 2^{3\!}P\) transitions are hence the dominant contributions to the polarisability over the frequencies considered, even though these transitions are far detuned, and the effect of the \(\MetastableState \rightarrow \UpperState\) transition is washed out due to it being highly forbidden.

The only other effect on the line shape due to the atomic polarisability is the Auter-Towns effect, which depends purely on the Rabi frequency \(\Omega\) \cite{PhysRevA.73.043416,CohenTannoudji1996}. The Rabi frequency is given by \(\Omega=\alpha E_{dc} \varepsilon_0\) \cite{PhysRevA.73.043416}, where \(\alpha\sim 2.45 \times 10^{-6}\)~Hz/(V/m)\(^2\) is the Stark polarisability, \(E_{dc}\) is the dc electric field amplitude (which can be constrained to be less than \(2\)~kV/m as described above), and \(\varepsilon_0\) is the ac electric field strength, which is \(\sim10^3\)~kV/m at the focus. We can hence constrain the Rabi frequency to be \(\Omega < 5\)~kHz, which is negligible compared to our linewidth, and hence we expect to see no Auter-Towns effect, which is what we observe.

\section{Details of Heating Method}


\begin{figure*}[t]
     \subfloat[\label{fig:tof}]{%
      \includegraphics[width=0.52\textwidth]{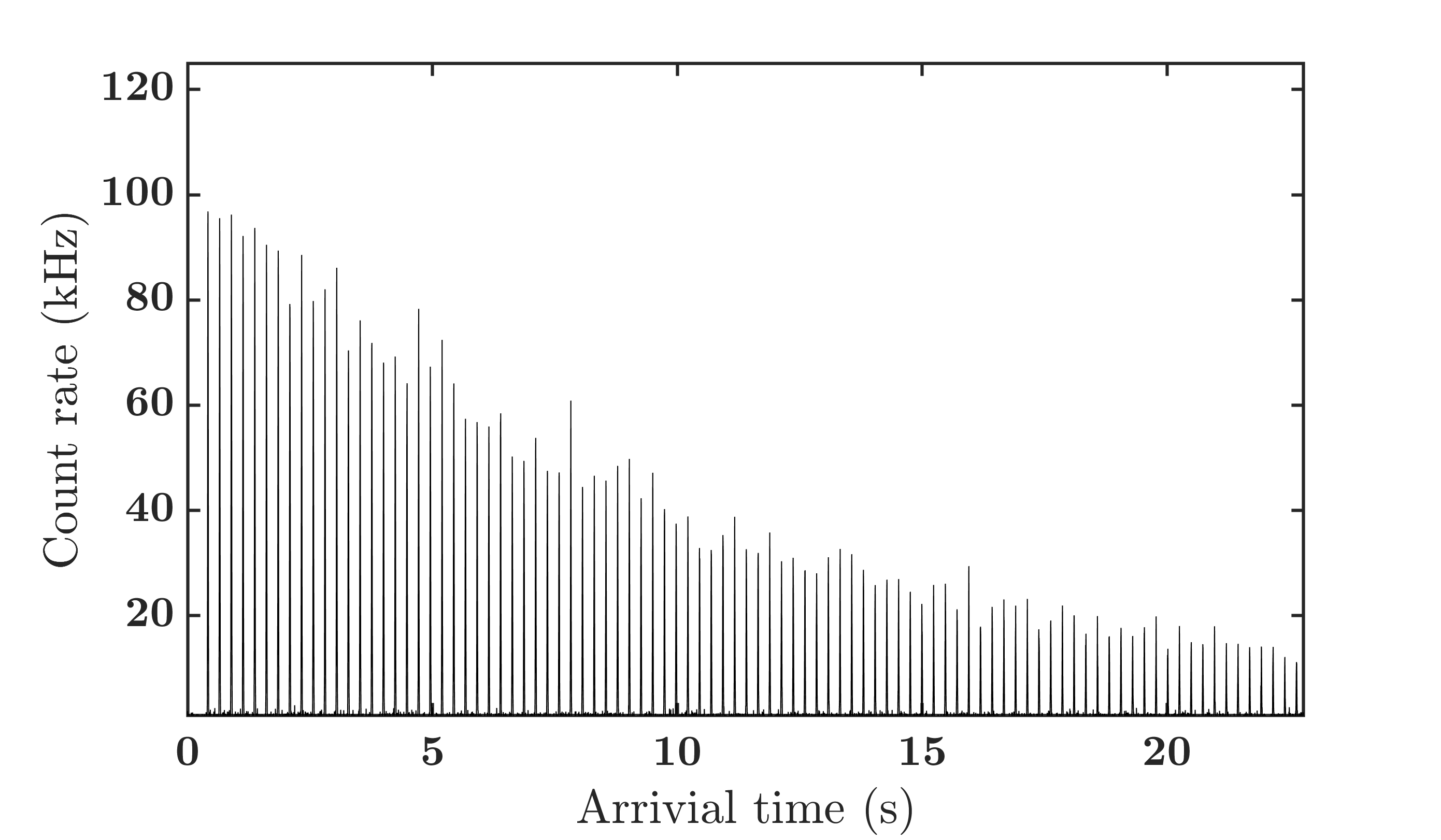}
     }
     \hfill
     \subfloat[\label{fig:pulse}]{%
      \includegraphics[width=0.42\textwidth]{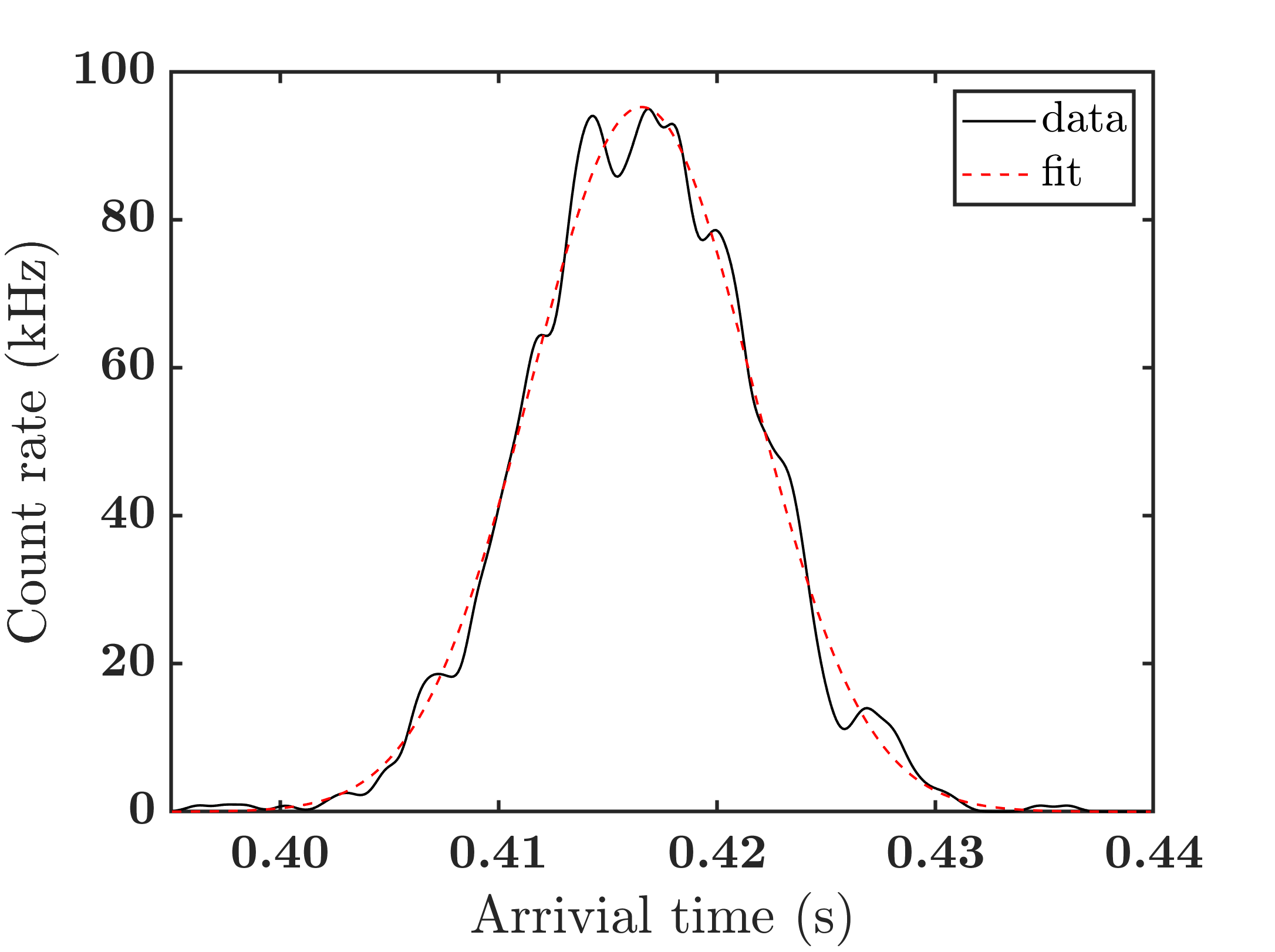}
     }
     \caption{(a) Time of flight profile for the heating method consisting of 95 outcoupled atomic pulses. The outcoupled atoms arrive at the detector in pulses, signified by the peaked structure of the profile. Arrival time is measured from when the first outcoupling pulse is applied. (b) Zoom in of the first pulse, with the solid black line representing count rate over time and the dashed red line indicating a Gaussian fit. This particular fit has parameters \(t_0=1.756(1)\)~s, \(\sigma_t = 0.0051(1)\)~s (equivalent to \(T=1.21(2)\)~\(\mu\)K), and \(C_0 = 96(2)\)~kHz.}
     \label{fig:dummy}
\end{figure*}


Fig.~\ref{fig:tof} displays the number of counts detected versus time, with a broadened radio frequency pulse applied to the trapped atoms every \(240\)~ms corresponding to the peaks present in the profile. If we zoom in on a particular peak (Fig.~\ref{fig:pulse}) we can see it has a distinctive Gaussian profile. 
From Yavin \textit{et al.} \cite{Yavin2002} we know that the expected time of flight probability density profile for a ballistic expansion of particles from a point source is
\begin{align}
    n(t) &= A \pi v_0^2 \left( \frac{\frac{1}{2}gt^2 + d }{t^2}\right) \exp \left( - \frac{(\frac{1}{2}gt^2 - d )^2}{v_0^2 t^2}\right), \label{eqn:tof_den_int}
\end{align}
where \(d\) is the fall distance from the trap to the detector, \(A = (m /2 \pi k_B T)^{3/2}\), \(m\) is the mass of a particle, \(k_B\) is the Boltzmann constant, \(v_0 = \sqrt{2 k_B T/m}\) is the most probable velocity, and \(g=-9.81\)~m/s\(^2\) is the acceleration due to gravity.If the spread of the peak is small in time we can simplify Eqn.~\ref{eqn:tof_den} by approximating it as a Gaussian
\begin{align}
    n(t) &\approx A \pi v_0^2 \left( \frac{1}{2}g + \frac{ d }{t_f^2}\right) \exp \left( - \frac{(t - t_f )^2}{ 2 \left(\frac{2}{g} \sqrt{\frac{k_B T}{m}}\right)^2}\right) \label{eqn:tof_den},
\end{align}
where \(t_f = \sqrt{\frac{2 d}{g}}\) is the expected fall time for particles with zero velocity.
Thus we fit a Gaussian to the count rate distribution in time \(C(t)\) of the form,
\begin{align}
    C(t) &= C_0 e^{-\frac{(t-t_0)^2}{2\sigma_t^2}},
\end{align}
where \(C_0\) is the peak count rate, \(t_0\) is the center time of the distribution (the time of arrival for particles with zero initial vertical velocity), and \(\sigma_t\) is the standard deviation. From Eqn.~\ref{eqn:tof_den} we have \(\sigma_t = \frac{2}{g} \sqrt{\frac{k_B T}{m}}\).
Rearranging we obtain,
\begin{align}
    T &= \left( -g \sigma_t/2\right)^2 \times \frac{m_{He}}{k_B}
\end{align}

\begin{figure}
    \centering
    \includegraphics[width=0.6\textwidth]{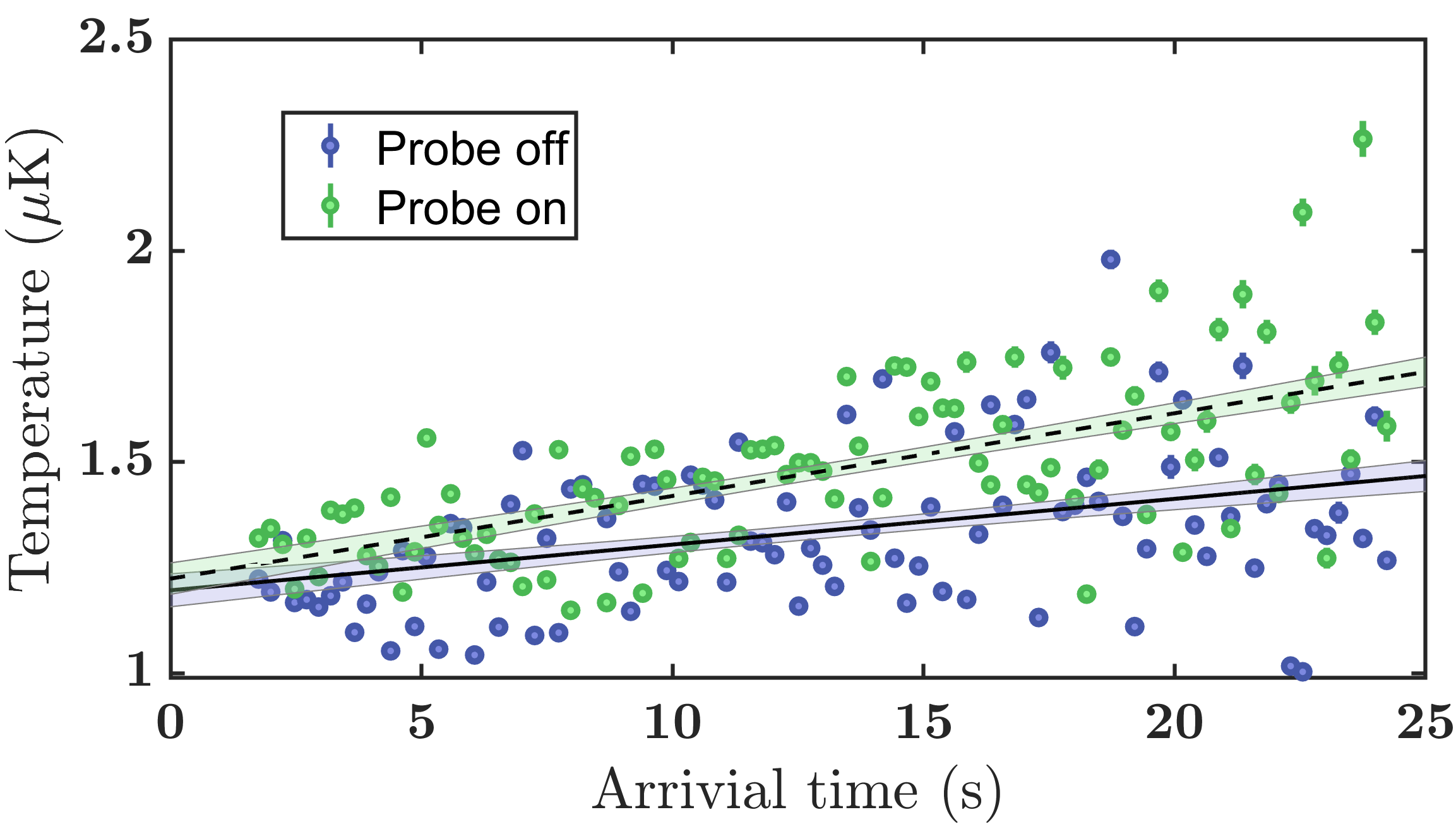}
    \caption{Measured temperature versus arrival time for a particular run of the heating method with the probe beam applied (dashed green) and the probe beam blocked as a reference (solid blue). Each individual point on the plot corresponds to an individual pulse, see Fig.~\ref{fig:tof} and Fig.~\ref{fig:pulse}. }
    \label{fig:heating_fig}
\end{figure}

From the fits of each pulse we extract the temperature versus time (see Fig.~\ref{fig:heating_fig}) and fit a line to this data, the gradient of which gives us the heating rate. The comparison of the heating rate with the probe applied to a reference then gives us a measure of the temperature increase purely due to photon absorption.

\section{Laser linewidth determination}
\label{sec:laser_linewidth}
To produce an independent estimation of the laser linewidth we employ spectroscopy of two cesium transitions. For determination of line-width up to $\sim$10~ks we use measurements of the two photon cesium $6^{2}S_{1/2} (F=4) \rightarrow 8^{2}S_{1/2} (F=4)$ transition at 364.5~THz (822.5~nm) which we detect using blue florescence in the cesium cell (the same as used in the wavemeter calibration as described in Sec.~\ref{sec:exp-details}) with a photonmultiplier tube (PMT) \cite{Fendel:07,Wu:13}. 

\begin{figure}
    \centering
    \includegraphics[width=0.6\textwidth]{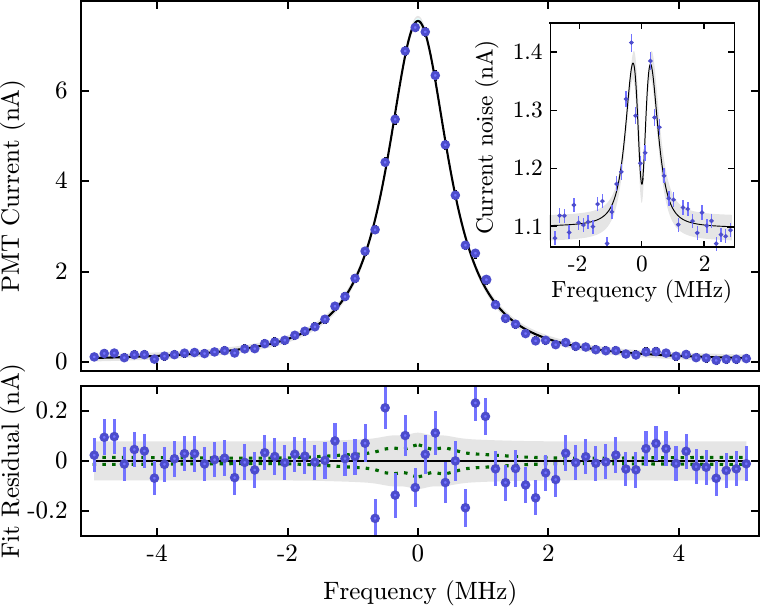}
    \caption{(Top) Example of a single scan across the two photon transition. Data points are marked in blue circles and Voigt fit is shown as the black line (error bars are smaller than maeker at this scale). Fit parameters are $\sigma=0.18(3)$~MHz $\gamma=0.49(3)$~MHz sentence (error is the standard deviation over many measurements). Inset shows fit to the PMT current noise with a given laser frequency noise component of \(0.075(1)\)~MHz. (Bottom) Residuals of the fit model, with shaded region shows one standard deviation of the observation error model and dashed lines are one standard deviation in the fit model. Current offset and fit frequency have been subtracted. Data shown here is acquired over \(75\)~s. }
    \label{fig:2p_scan_single}
\end{figure}

We acquire data by adjusting the set-point of our wavemeter feedback system and measuring the PMT current produced. We fit the observed transition with a Voigt profile, a convolution of a Gaussian with standard deviation \(\sigma\) and Lorentzian with  scale parameter \(\gamma\), as shown in Fig.~\ref{fig:2p_scan_single} with fit parameters $\sigma=0.18(3)$~MHz and $\gamma=0.49(3)$~MHz sentence (error is the standard deviation over many measurements). Note that the Lorentzian FWHM is within error of the predicted value of $\gamma=0.48$~MHz corresponding to the combined effect finite upper state lifetime ($\gamma=0.46$~MHz) and transit time broadening \cite{Fendel:07}. From this we can constrain the laser linewidth over timescales of the scans ($\sim$70~s) to be $\sigma\approx0.18(3)$~MHz and $\gamma<0.03$~MHz corresponding to the Gaussian and Lorentzian components of the un-doubled (red) probe laser system.
 
It is also possible to obtain an independent measurement of the laser frequency noise by using the noise in the PMT current as a function of the frequency. A strong component of this noise is from the transduction of laser frequency noise through the derivative of the line profile into the PMT current noise, which we define as the standard deviation of the measured PMT current. We use a model which combines this mechanism along with shot noise and background terms (in quadrature). The amplitude and center frequency used in these terms are fixed from the fit to the mean current. The result of one such fit is shown in the inset of Fig.~\ref{fig:2p_scan_single}. From this we get a laser frequency noise of $0.07(1)$~MHz between 2~Hz and 3.5~kHz. 

To find the contribution to the laser linewidth from drifts at timescales greater than the scan duration we introduce an estimator of the standard deviation \(\sigma^{2}(f(t),T,\tau)\) over integration duration $\tau$, which uses the center frequency \(\mu\) from the fits to multiple scans over about 10~h,
\begin{align}
    \sigma^{2}(f(t),T,\tau)&=\frac{1}{\tau}\int_{T}^{T+\tau}   (f(t)-\mu(f(t),T,\tau))^2 dt \\ 
    \mu(f(t),T,\tau)&=\frac{1}{\tau}\int_{T}^{T+\tau}  f(t) dt
\end{align}

\begin{figure}
    \centering
    \includegraphics[width=0.6\textwidth]{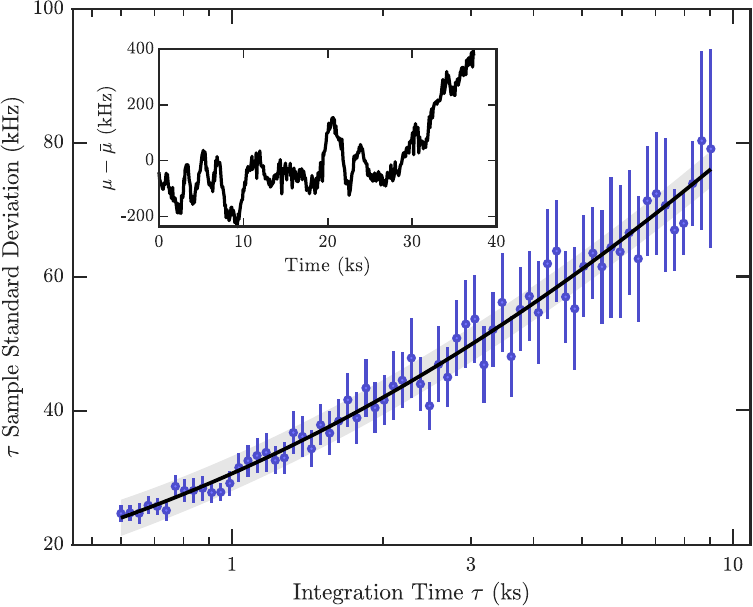}
    \caption{Dependence of the standard deviation of the scan fit center frequency ($\mu$) with integration time $\tau$. Fit is a second order polynomial in log space $\sigma_f(\tau)=o+g\log_{10}{(t/1s)}+c\log_{10}{(t/1s)}^2$ with $o=66$~kHz, $g=-60(20) \mathrm{kHz}$, and $c=15(3) \mathrm{kHz}$. Inset shows the values of $\mu$ about the mean.}
    \label{fig:2p_tau_sd_dep}
\end{figure}

From Fig.~\ref{fig:2p_tau_sd_dep} we see a monotonic increase in the variation in the laser frequency. We cannot extrapolate from 10~h that these scans were taken over, to the 12 days that the main transition data was taken over. To this end we use two separate methods to estimate the effective laser linewidth at these time scales. 

In order to support the accuracy of this measurement we can use the cesium (SAS) calibration data as seen in Fig.~\ref{fig:wm_model}, and described in Sec.~\ref{sec:exp-details}. The standard deviation of this data is \(1.7(5)\)~MHz. To obtain the laser linewidth we combine in quadrature with the Gaussian component of the the short term laser linewidth \(0.36(6)\)~MHz (converted to the blue) giving \(1.7(5)\)~MHz, which is within error of the Voigt predicted Gaussian linewith for both sets of data. Here we have neglected the Lorentzian contribution which we have previously constrained to be $\gamma<0.06$~MHz.

For are main method we perform a Voigt fit directly to the data which gives a Guassian component of \(\sigma=1.9(4)\) for the direct detection method and \(\sigma=1.6(9)\) for the heating method.




\section{Theoretical Value of the excited state lifetime of the \(\UpperState\) state in helium}
The excited state lifetime is the average amount of time an atom will remain in a particular excited state before decaying to a lower energy state. The state lifetime \(\tau_u\) can be calculated for a given state \(u\) from the Einstein \(A\) coefficients of all transitions to lower lying states,
\begin{align}
    \tau_u &= \frac{1}{\sum_l A_{ul}},
\end{align}
where \(A_{ul}\) denotes the Einstein \(A\) coefficient for the transition between the upper \(u\) and the lower state \(l\). For the \(\UpperState\) state of helium the major contributions to the state lifetime are from the transitions to the \(\MidState\) states, which have respective Einstein \(A\) coefficients \(J=0\) \(A = 3.095(9)\times 10^{6}\)~\(\text{s}^{\text{-}1}\), \(J=1\) \(A = 9.28(3)\times 10^{6}\)~\(\text{s}^{\text{-}1}\), and \(J=2\) \(A = 1.55(5)\times 10^{7}\)~\(\text{s}^{\text{-}1}\) \cite{NISTASD}. Hence the theoretically expected state lifetime of the \(\UpperState\) state is \(\tau = 35.9(2)\)~ns.

\section{Experimental determination of Excited state lifetime}

The scattering probability distribution in frequency space of an individual transition has a fundamental limit of a Lorentzian distribution, whose linewidth, or more precisely full width half maximum (FWHM), \(\Gamma\) is related to the state lifetime of the excited state of the transition \(\tau\) by \(\tau = 1/(2\pi \Gamma)\). 

Due primarily to the finite laser linewidth, and Gaussian profile, of the probe laser the measured signal of the transition takes the form of a Voigt distribution. In order to extract the underlying Lorentzian profile of the transition we fit a Voigt profile to the measured data (see Fig.~\ref{fig:427nm_signal} and Fig.~\ref{fig:427nm_signal_heating} respectively in main text). We ensure that the Gaussian component of the Voigt profile is largely produced by the laser line profile as described in Sec.\ref{sec:laser_linewidth}. The Voigt profile hence directly gives the Lorentzian linewidth, which we find to be \(\Gamma_d = 3.2(10)\)~MHz for the direct detection method, and \(\Gamma_h=4(3)\)~MHz for the heating method, with the uncertainties given by the confidence interval of the respective fits. This gives us our value for the excited state (\(\UpperState\)) lifetime of \(\tau = 50(16)\)~ns for the direct detection method and \(\tau = 40(30)\)~ns for the heating method.

The other primary source of broadening is the mean field effect, however this is negligible compared to the laser linewidth. An approximate value for the mean field broadening can be obtained from the mean field shift calculated in Sec.~\ref{sec:freq_shifts}, hence for both methods we have \(\sigma_{mf} \sim 0.01\)~MHz, which converted to a FWHM is \(\Gamma_{mf} \sim 0.024\)~MHz, far below that of the laser linewidth.

\section{Experimental determination of Einstein \(A\) coefficient}



In this section we present a derivation of an expression for the Einstein \(A\) coefficient that is entirely dependant on empirical data and known quantities, starting from a fundamental expression, Eqn.~\ref{eqn:A_main}. First consider light with frequency \(f\) and intensity distribution \(I_{f}(x,y,z)\) propagating along the \(x\)-axis through some absorbent medium. The change in intensity at a particular point \(\delta I_{f} (x,y,z)\) over a small distance \(\delta x\) is given by the expression \cite{Huber1986}
\begin{align}
    - \delta I_{f} (x,y,z) &= k(f,x,y,z) I_{f} (x,y,z) \delta x \label{eqn:A_main},
\end{align}
where \(k(f,x,y,z)\) is the frequency and density dependant absorption coefficient. Let the total power in the light field at point \(x\) be \(P_{f}(x)\) and \(I_{f} (x,y,z) = P_{f}(x) \phi_I(x,y,z)\), where \(\phi_I(x,y,z)\) is a function that gives the distribution of the power over a given plane and has unit normalisation. We can hence integrate Eqn.~\ref{eqn:A_main} over the \(y\)-\(z\) plane, the plane perpendicular to the direction of light propagation, and obtain the change in power \(\delta P_f(x)\) over an infinitesimal distance \(\delta x\) is
\begin{align}
    - \delta P_{f}(x) &= P_{f}(x) \delta x \int \int  dy dz  \, k(f,x,y,z) \phi_I(x,y,z)\\
    \therefore - \frac{\delta P_{f}(x)}{P_{f}(x)} &= \delta x \int \int  dy dz \, k(f,x,y,z) \phi_I(x,y,z). \label{eqn:temp_1}
\end{align}
Both sides of Eqn.~\ref{eqn:temp_1} can be integrated to obtain,
\begin{align}
    - \int \frac{d P_{f}(x)}{P_{f}(x)} &= \int \int \int dx  dy dz \, k(f,x,y,z) \phi_I(x,y,z)\\
    -\log\left( \frac{P_{f}^+}{P_{f}^-} \right) &= \iiint_{all\, space} \, k(f,x,y,z) \phi_I(x,y,z) \label{eqn:temp_2},
\end{align}
where \(P_{f}^-\) is the power in the light field before moving through the medium, in a theoretical sense the power at \(x=-\infty\), and \(P_{f}^+\) is the power in the light field after the medium, \textit{i.e.} the power at \(x=+\infty\). Next we note that the integral of the absorption coefficient over the line shape of a transition is related to the Einstein \(A\) coefficient as follows \cite{Huber1986}:
\begin{align}
    \int_{line} k(f,x,y,z) df &= \frac{c^2}{8 \pi f_0^2} A_{ul} \, n(x,y,z),
\end{align}
where \(f_0\) is the center frequency of the transition, \(A_{ul}\) is the Einstein \(A\) coefficient for the transition between the states \(u\) and \(l\), and \(n(x,y,z)\) is the atomic density distribution at the point \((x,y,z)\). Therefore Eqn.~\ref{eqn:temp_2} can be written as
\begin{align}
   -\int df \log\left( \frac{P_{f}^+}{P_{f}^-} \right) &= \iiint_{all\, space} \, \frac{c^2}{8 \pi f_0^2} A_{ul} \, n(x,y,z) \, \phi_I(x,y,z) \label{eqn:temp_3},
\end{align}
where we have assumed the intensity distribution is frequency independent. To simplify Eqn.~\ref{eqn:temp_3} further, we assume that the power removed from the beam is small and so we can Taylor expand \(-\log\left( \frac{P_{f}^+}{P_{f}^-} \right)\) about \(1\) as \(-\log\left( \frac{P_{f}^+}{P_{f}^-} \right) \approx \frac{P_{f}^--P_{f}^+}{P_{f}^-}\). The difference between the initial and final powers is equal to the rate of photons scattered by the atoms multiplied by the energy of the photon, \(P_{f}^--P_{f}^+= h f \frac{dN_{scatter}}{dt}\). The number of scattered photons can be calculated from the frequency dependant heating rate due to the probe beam \(\frac{dT}{dt}(f)\), the heat capacity of the atoms, which for a thermal gas in a constant harmonic trap is \(C_b = 3 N k_B\) \cite{doi:10.1119/1.3417868}, and the average energy added to the gas by each photon \(E_p\), from the relation \(\frac{dN_{scatter}}{dt} = \frac{dT}{dt}(f) \frac{C_b}{E_p}\). We can simplify even further by utilising the relation \(n(x,y,z) = N \phi_N(x,y,z)\), where \(N\) is the total atom number and \(\phi_N(x,y,z)\) gives the atomic density distribution with unit normalisation. Combining we see Eqn.~\ref{eqn:temp_3} becomes,
\begin{align}
    \int df \frac{hf}{P_{f}^-} \frac{dT}{dt}(f) \frac{C_b}{E_p} &= \frac{c^2}{8 \pi f_0^2} A_{ul} \iiint_{all\, space} \,  \, \phi_N(x,y,z) \, \phi_I(x,y,z) \\
    \therefore A_{ul} &= \frac{24 \pi f_0^2 h k_B}{c^2 E_p} \frac{ \int df f \left(P_{f}^-\right)^{-1} \frac{dT}{dt}(f)}{\iiint_{all\, space} \,  \, \phi_N(x,y,z) \, \phi_I(x,y,z)}. \label{eqn:A_final}
\end{align}
The parameters in Eqn.~\ref{eqn:A_final} are all experimentally measured or determined as follows: \(f\) is measured by the HighFinness wavemeter, \(P_{f}^-\) is measured by a calibrated photodiode, \(\frac{dT}{dt}\) is extracted from the data as described in the heating method section, \(\phi_I(x,y,z)\) is measured via a camera, \(\phi_N(x,y,z)\) can be determined by the trapping frequency and temperature \cite{doi:10.1119/1.3417868}, \(E_p\) is the energy transferred to the cloud on average by the recoil momentum of the photons, and the remaining parameters are all known constants.

\begin{figure}
    \centering
    \includegraphics{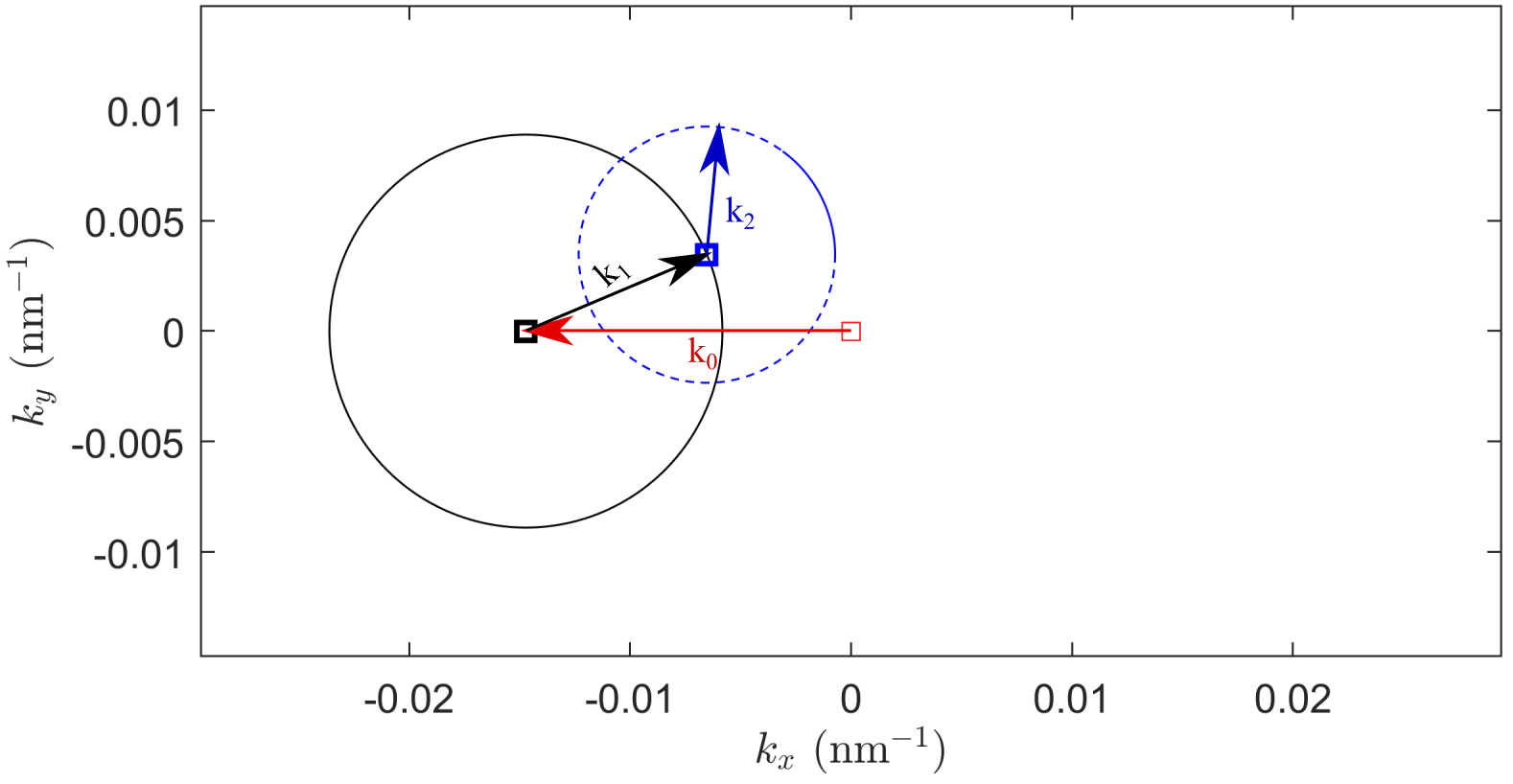}
    \caption{Cross section of the momentum distribution (in wavenumber space) of an atom after it has under gone the \(\MetastableState \rightarrow \UpperState\) (\(k_0\)), \(\UpperState \rightarrow \MidState\) (\(k_1\)), and \(\MidState \rightarrow \MetastableState\) (\(k_2\)) transitions. Note that three arrows show are meant to be representative of a particular possible outcome, and the circles are crosssection representations of spheres.}
    \label{fig:k_dist}
\end{figure}

The energy transferred to the cloud from photon recoils can be estimated by \(E_p = \eta \frac{1}{2m} \left(\frac{h}{c}\right)^2(f_0^2 + f_1^2 + f_2^2)\), where \(\frac{1}{2m} \left(\frac{h f}{c}\right)^2\) is the energy of a single photon recoil for a given photon of frequency \(f\), \(f_0\), \(f_1\), and \(f_2\) are the frequencies for the the \(\MetastableState \rightarrow \UpperState\), \(\UpperState \rightarrow \MidState\), and \(\MidState \rightarrow \MetastableState\) transitions respectively, and \(\eta\) is the probability that an excited photon transfers its recoil energy to the cloud. Note that we can make the approximation that all excitations have the same photon recoil as the linewidth of the transition is negligible compared to its center frequency. To see that the average energy transferred to the atom from the three transitions is the sum of their respective photon recoils consider the momentum distribution after absorption and the two emissions, see Fig.~\ref{fig:k_dist}, the average momentum is hence given by the surface integral
\begin{align*}
    \avg{k^2} &= \frac{1}{4\pi k_1^2 \times 4\pi k_2^2}\int^{2\pi}_0 d \theta_1 \int^{\pi}_0 d \phi_1 \int^{2\pi}_0 d \theta_2 \int^{\pi}_0 d \phi_2 \, [\left(k_0 + k_1 \sin(\theta_1)\cos(\phi_1) + k_2 \sin(\theta_2)\cos(\phi_2)\right)^2+...\\
     &\, \, \, \, \, \, \, \, \left(k_1 \sin(\theta_1)\sin(\phi_1) + k_2 \sin(\theta_2)\sin(\phi_2)\right)^2+\left(k_1\cos(\phi_1) + k_2 \cos(\phi_2)\right)^2]k_1^2 \sin(\phi_1) k_2^2 \sin(\phi_2)\\
    &= k_0^2 + k_1^2 + k_2^2
\end{align*}
where \(k_i = 2\pi f_i/c\), is the wavenumbers for the respective transition. 

To calculate \(\eta\) we assume that all atoms which decay to a trapped state, \(24.07\%\) of excited atoms, thermalise with the cloud, and those that decay to an untrapped state have an additional probability of colliding with another atom in the BEC or cloud as the excited atom leaves. To determine this additional probability we perform a Monte Carlo simulation of an atom leaving the cloud. The procedure for this simulation is as follows, we randomly sample from the momentum distribution of the cloud, given by the trapping parameters and temperature, and from the momentum distribution produced by the three photon recoils due to the absorption from \(\MetastableState\) to the \(\UpperState\) excited state and then the decays to the \(\MidState\) and then \(\MetastableState\) states, see Fig.~\ref{fig:k_dist}. These momenta are then added together and the atoms initial position is also sampled from the distribution given by the temperature and trapping parameters. The atom is then propagated ballisticly under gravity. We then simulate the probability of the atoms colliding with another atom at a specific point using the atomic density distribution and the scattering cross section \(\sigma = 4 \pi a^2\) where \(a\) is the \(s\)-wave scattering length of the \(\MetastableState\) state. We then repeat this process many times to find the average probability of a collision, and hence thermalisation, with the cloud, which we denote \(\eta_c\). The total probability is hence \(\eta = 24.07\% + \eta_c\).

\textbf{Error propagation for the Einstein \(A\) coefficient:} 
Eqn.~\ref{eqn:A_final} can be simplified as \(A=\frac{C I_1}{E_p I_2}\), with \(C=\frac{24 \pi f_0^2 h k_B}{c^2}\), \(E_p\) as defined above, \(I_1 = \int df f \left(P_{f}^-\right)^{-1} \frac{dT}{dt}(f)\) and \(I_2 = \iiint_{all\, space} \,  \, \phi_N(x,y,z) \, \phi_I(x,y,z)\).
Using regular propagation of uncertainty we find the uncertainty in \(A\), which we denote \(\delta A\), to be
\begin{align}
    \delta A &= A \sqrt{\left(\frac{\delta C}{C}\right)^2 + \left(\frac{\delta I_1}{I_1}\right)^2 + \left(\frac{\delta E_p}{E_p}\right)^2 + \left(\frac{\delta I_2}{I_2}\right)^2 }
\end{align}
Since the error in \(C\) is negligible we have \(\delta C =0\). We can propagate the error in \(I_1\) and \(I_2\) using bootstrapping techniques. The error in \(E_p\) is dominated by uncertainty in the Monte Carlo simulation, which can be estimated through the standard deviation of the output. With the above the error in \(A\) can hence be calculated.

\section{Clebsch-Gordan coefficients}
The Clebsch-Gordan coefficients represent the angular momentum coupling between different atomic states. For our work they are relevant because the modulus squared of the  Clebsch-Gordan coefficient between two particular magnetic substates is equal to the relative intensity, or probability, of that transition in comparison to all other transitions between the two respective manifolds \cite{Krainov2019}. The relative transition strengths, normalised to the weakest allowed transition, between the \(^{3\!}S_1\) and \(^{3\!}P_{0,1,2}\) manifolds are given in Tab.~\ref{tab:clebsch_gordan} \cite{metcalf1999laser}. Given that the excited particle is initially in the \(\UpperState, \, m_J=0\) state and the transitions from this state are dominated by decays to the \(\MidState\) states, and then from those states the transitions are dominantly towards the \(\MetastableState\) state, we can calculate the probability of an atom decaying to each magnetic sub state from Tab.~\ref{tab:clebsch_gordan}. The initial relative transition probabilities from the \(\UpperState, \, m_J=0\) state are given by the second row of Tab.~\ref{tab:clebsch_gordan}, and the exact fraction can be obtained by normalising the row by its sum \(18\). The transitions from the \(\MidState\) magnetic sub states to the \(\MetastableState, \, m_J=(+1,0,-1)\) sub states are given by the column of each relevant state (specifically all states except \(2^{3\!}P_{2}, \, m_J=\pm2\) as \(m_J\) can at most change by one), and again the fractions can be obtained by normalising each column by its total \(6\). From this we obtain the fraction of atoms that decay down each of the possible paths, and after summing the fractions which lead to each final state we obtain the total fraction that end up in each \(\MetastableState, \, m_J=(+1,0,-1)\) state as \(\left(\frac{26}{108},\frac{56}{108},\frac{26}{108}\right)\) or as approximate percentages \(\left(24\%,52\%,24\%\right)\).
\begin{table}[]
    \centering
    \begin{tabular}{@{\extracolsep{\fill}}>{\centering}p{0.06\textwidth} |
    @{\extracolsep{\fill}}>{\centering}p{0.06\textwidth} |
    @{\extracolsep{\fill}}>{\centering}p{0.07\textwidth} |
    @{\extracolsep{\fill}}>{\centering}p{0.07\textwidth}
    @{\extracolsep{\fill}}>{\centering}p{0.07\textwidth}
    @{\extracolsep{\fill}}>{\centering}p{0.07\textwidth}|
    @{\extracolsep{\fill}}>{\centering}p{0.07\textwidth}
    @{\extracolsep{\fill}}>{\centering}p{0.07\textwidth}
    @{\extracolsep{\fill}}>{\centering}p{0.07\textwidth}
    @{\extracolsep{\fill}}>{\centering}p{0.07\textwidth}
    @{\extracolsep{\fill}}>{\centering\arraybackslash}p{0.07\textwidth}}
    \toprule
    \toprule
         \(^{3\!}S_1\) &\multicolumn{10}{c}{\(^{3\!}P_J\)}  \\
         \hline
         \(m_J\)& J & \multicolumn{1}{c|}{\!0} & \multicolumn{3}{c|}{\!\(1\)} & \multicolumn{5}{c}{\!2} \\
         & \(m_J\)&0&\(+1\)&0&\(-1\)&\(+2\)&\(+1\)&\(0\)&\(-1\)&\(-2\)\\
         \hline
         \(+1\)& &\(2\)&\(3\)&\(3\)&\(-\)&\(6\)&\(3\)&\(1\)&\(-\)&\(-\) \\
         \(0\)& &\(2\)&\(3\)&\(0\)&\(3\)&\(-\)&\(3\)&\(4\)&\(3\)&\(-\) \\
         \(-1\)& &\(2\)&\(-\)&\(3\)&\(3\)&\(-\)&\(-\)&\(1\)&\(3\)&\(6\)\\
    \bottomrule
    \bottomrule
    \end{tabular}
    \caption{Transition strengths in the D-line of He* obtained from the mod square of the  Clebsch-Gordan coefficient and normalised to the weakest allowed transition \cite{metcalf1999laser}. To obtain fractional transition rates divide the values in the relevant row or column by its respective total (\(18\) for rows and \(6\) for columns).}
    \label{tab:clebsch_gordan}
\end{table}

\section{RF Knife}

For the direction detection method a constant radio frequency field was applied to the atoms during the detection phase. The reason for this was to attempt to increase the detection efficiency of atoms which absorb a photon by both outcoupling some portion of atoms which decay back to the trapped \(\MetastableState,\, m_J=+1\) state, and lensing these atoms so that a higher proportion would land on the detector.

Consider that the two subsequent photon decays produce a momentum distribution which is a filled shell with a center translated by an outer radius equal to a 427~nm photon recoil. As the maximum magnetic field experienced by an atom oscillating in the magnetic trap is proportional to its maximum kinetic energy, by applying RF radiation tuned to the \(\MetastableState, m_J=1 \rightarrow 0\) transition for a given magnetic field strength we can selectively outcouple atoms with maximum kinetic energy above a threshold. Furthermore, as the trap does work on an atom the momentum space distribution of these outcoupled atoms is reduced in size, hence improving collection efficiency. Thus while only $\sim$24\% of atoms decay to the \(\MetastableState,\, m_J=+1\) state via this process they could theoretically significantly increase the total signal amplitude. The interplay between these two effects produce an optimum collection efficiency that depends on the particular momentum distribution and detector geometry.

It was found experimentally, however, that there was no statistically significant increase in the signal amplitude between the RF field applied and not applied. The exact reasons for this are unknown but we conjecture that the atoms scattering as they leave the trap and the RF not saturating the transition are the most likely contributing factors.

\section{Sensitivity Metric}

We can define a metric in order to quantify the sensitivity of our measurement techniques in terms of minimum detectable Einstein A values. Our method is expected to have a signal to noise ratio (SNR) which is given by Poissionian statistics as,
\begin{equation}
    \mathrm{SNR}= \frac{A \sqrt{N t P \eta }}{S}  ,
\end{equation}
where $A$ is the Einstein \(A\) Coefficient, $N$ is the atom number, $\eta$ is the detector QE, $t$ is the integration time, $P$ is the beam power. The sensitivity $S$ is defined as the transition with the smallest Einstein \(A\) Coefficient ( \(A_{min}\)) that can be detected per square root of the previous signal scaling paramters
For the direct detection method we have \(\mathrm{SNR} = 26\) (for a single dataset ), \(N=1.7\times 10^6\) atoms, \(t=25\times215\)~s (as we interrogate for \(25\)~s for each of the total of \(215\) measurement shots), \(P=26\)~mW, and $\eta=0.09$. 
Thus we find the specific sensitivity to be \(S\approx 1\times10^{-6}\)~\(\text{s}^{-1} \sqrt{s\cdot W}\) and the minimum detectable value ($\mathrm{SNR}=1$) for our experiment at \(A_{\mathrm{min}}^{\mathrm{direct}} \approx 7 \times10^{-11}\)~\(\text{s}^{\text{-}1}\) (\(1/A \approx 470\)~years)  for a combined day of interrogation. 
The heating method is less sensitive with \(\mathrm{SNR} = 8\),  \(N=1\times 10^7\) atoms, \(t=25\times188\)~s, and
\(P=17.8\)~mW, giving \(S\approx 8\times10^{-6}\)~\(\text{s}^{\text{-}1} \sqrt{s\cdot W}\). The corresponding minimum detectable value ($\mathrm{SNR}=1$) for this method is \(A_{\mathrm{min}}^{\mathrm{heat}} \approx 2 \times10^{-10}\)~\(\text{s}^{\text{-}1}\) (\(1/A \approx 160\)~years) for a combined day of interrogation. 
Both methods are currently limited by technical effects, the direct detection by the by the dark count rate of the detector system \(\sim130/s\) and the heating rate by the background heating rate of trapped atoms. These current limits could be improved on by using MCPs with lower radioisotope levels for lower background count rate and using a weaker trap where the heating rate is lower.

\end{document}